\documentclass[10pt,amssymb,superscriptaddress,showkeys,twocolumn,prl]{revtex4-2}
\usepackage{graphicx}
\usepackage{dcolumn}
\usepackage{bm}
\usepackage{changes}
\usepackage{amsmath,amssymb}
\begin{document}
\title{Octupole topological insulating phase in Brillouin three-dimensional real projective space}

\author{Sichang Qiu}
\thanks{These authors contributed equally}
\affiliation{State Key Laboratory of Millimeter Waves, Southeast University, Nanjing 210096, China.}
\author{Jinbing Hu}
\thanks{These authors contributed equally}
\affiliation{College of Optical-Electrical Information and Computer Engineering, University of Shanghai for Science and Technology, Shanghai 200093, China}
\author{Yi Yang}
\affiliation{Department of Physics and HK Institute of Quantum Science and Technology, The University of Hong Kong, Pokfulam, Hong Kong, China}
\author{Ce Shang}
\email{shangce@aircas.ac.cn}
\affiliation{Aerospace Information Research Institute, Chinese Academy of Sciences, Beijing 100094, China}
\author{Shuo Liu}
\email{liushuo.china@seu.edu.cn}
\affiliation{State Key Laboratory of Millimeter Waves, Southeast University, Nanjing 210096, China.}
\author{Tie Jun Cui}
\email{tjcui@seu.edu.cn}
\affiliation{State Key Laboratory of Millimeter Waves, Southeast University, Nanjing 210096, China.}
\date{\today}

\begin{abstract}

Recent advancements in quantum polarization theory have propelled the exploration of topological insulators (TIs) into the realm of higher-order systems, leading to the study of the celebrated two-dimensional (2D) quadrupole and three-dimensional (3D) octupole TIs. Traditionally, these topological phases have been associated with the toroidal topology of the conventional Brillouin zone (BZ). This Letter reports on the discovery of a novel octupole topological insulating phase emerging within the framework of the Brillouin 3D real projective space (\( \mathbb{RP}^3 \)). We theoretically propose the model and its corresponding topological invariant, experimentally construct this insulator within a topological circuit framework, and capture the octupole insulating phase as a localized impedance peak at the circuit's corner. Furthermore, our \( \mathbb{RP}^3 \) circuit stands out as a pioneering 3D model to simultaneously exhibit both intrinsic, termination-independent symmetry-protected topological phases (SPTPs) and extrinsic, termination-dependent boundary-obstructed topological phases (BOTPs), which broadly encompass 2D surface-obstructed topological phases (SOTPs) and 1D hinge-obstructed topological phases (HOTPs). Our results broaden the topological landscape and provide insights into the band theory within the manifold of the Brillouin $\mathbb{RP}^3$.

\end{abstract}
            
                            \maketitle

Topological insulators (TIs), remarkable for their robustness against internal defects and external perturbations, have risen to a research prominence in various areas such as photonics \cite{PhysRevB.98.201114,PhysRevB.98.205147,PhysRevLett.111.243905,PhysRevLett.114.037402,PhysRevLett.120.063902}, acoustics \cite{PhysRevLett.114.114301,PhysRevB.96.094106,PhysRevB.96.184305,PhysRevLett.122.244301,PhysRevLett.124.206601}, mechanics \cite{PhysRevLett.116.135503,PhysRevApplied.11.044029}, and electrical circuits \cite{PhysRevB.98.201402,RN1,PhysRevB.99.121411,PhysRevResearch.2.023265,PhysRevLett.125.166801,PhysRevLett.126.146802,RN2,Liu_2021,PhysRevApplied.13.014047}. The field of topological materials has witnessed significant advancements, ranging from first-order systems to higher-order topological insulators (HOTIs). HOTIs transcend the conventional bulk-edge correspondence theory,  featuring boundary states in dimensions lower than $n-1$ \cite{PhysRevLett.124.036803,PhysRevB.98.201114,PhysRevLett.122.204301,RN3,RN16,RN17,PhysRevLett.124.136407,PhysRevLett.125.146401,PhysRevLett.125.146401,PhysRevLett.122.256402,PhysRevB.98.205147}. To date, the topological properties of the aforementioned research are based on the Brillouin zone (BZ) torus $\mathbb{T}^n$  ( $=\underbrace {{\mathbb{S}^1} \times {\mathbb{S}^1} \times  \cdots  \times {\mathbb{S}^1}}_n$ is an orientable $n$-dimensional manifold defined as the product of the bundle of $\mathbb{S}^1$ cylinders), where the Bloch hamiltonian $H({\bold{h}})$ is restricted to the first Brillouin zone and defined with a reciprocal lattice vector $\bold G$ as $H({\bold{h}})=H({\bold{h+G}})$ \cite{PhysRevB.90.165114,RevModPhys.88.035005,PhysRevB.90.205136,PhysRevX.9.041015,PhysRevLett.121.126402}.

However, the torus is not the only example of a closed compact manifold; the Klein bottle and the real projective plane also belong to this category. Under the \(\mathbb{Z}_2\) gauge field \cite{PhysRevLett.118.070501,PhysRevLett.98.087204,PhysRevLett.124.120503,PhysRevLett.105.080501,PhysRevLett.128.116802,PhysRevLett.129.043902} with the alternative signs of the hopping amplitudes, the symmetries of the system would satisfy projective algebra, extending the Bloch band theory based on $\mathbb{T}^2$ BZ to Klein ${\mathbb{K}^2}$ ($={\mathbb{S}^1} \times {\mathbb{X}^1}$ with ${\mathbb{X}^1}$  defining  M\"{o}bius bundle) BZ manifold \cite{RN8,PhysRevB.108.235412}. Specifically, the projective symmetry algebra generates an unconventional `momentum-space non-symmorphic symmetry (\(\mathbf{k}\)-NS)', which contains a fractional translation in the reciprocal lattice. Such phenomena have already been experimentally demonstrated in acoustic crystals in the form of M\"{o}bius insulators \cite{PhysRevLett.128.116802, PhysRevLett.128.116803}. Recent studies have shown that a real projective plane $\mathbb{RP}^2$ ($={\mathbb{X}^1} \times {\mathbb{X}^1}$) BZ can be employed to construct two-dimensional (2D) HOTIs with quadrupole moments \cite{PhysRevB.108.235412,Hu2023HigherOrderTI}.  Research has also recently been conducted on the development of HOTIs within the \(\mathbb{RP}^2\) of the real space \cite{RN5}. 
The concept is also associated with the Brillouin Klein space $\mathbb{K}^3$ ($={\mathbb{X}^1} \times {\mathbb{S}^1}\times {\mathbb{S}^1}$) or the half-turn space $\mathbb{HT}^3$ ($={\mathbb{X}^1} \times {\mathbb{X}^1}\times {\mathbb{S}^1}$) \cite{RN35}. The comprehensive understanding of fundamental theory remains incomplete, with the three-dimensional real projective space $\mathbb{RP}^3$ ($={\mathbb{X}^1} \times {\mathbb{X}^1}\times {\mathbb{X}^1}$) representing the elusive final piece of the puzzle in 3D. Moreover, there are two different classifications of HOTIs \cite{PhysRevB.97.205135,PhysRevResearch.3.013239}: intrinsic HOTIs, which host symmetry-protected topological phases (SPTPs) induced by bulk gap closures, and extrinsic HOTIs, which host boundary-obstructed topological phases (BOTPs) dependent on boundary termination. To date, no 3D HOTI that simultaneously involves both \(\mathbf{k}\)-NS symmetries and the coexistence of SPTPs and BOTPs features has been reported.

In this paper, we propose a 3D HOTI in Brillouin \( \mathbb{RP}^3 \), which hosts higher-order corner states induced by the octupole moment of the bulk. Unlike the BBH model \cite{RN12,RN3}, which also hosts bulk octupole moment, we introduce \(\mathbf{k}\)-NS symmetries along all three axes in momentum space by enforcing the \(\mathbb{Z}_2\) gauge field with a chessboard $\pi$-flux configuration to the 3D lattice,  transforming the original BZ as a manifold with three pairs of opposing faces glued by a half-twist method \cite{RN8,Hu2023HigherOrderTI}. These unconventional symmetries divide the 3D BZ into 64 blocks, grouped into eight categories, where any set of eight blocks, one from each category, forms a reduced BZ that retains all the essential information of the original BZ \cite{suppmat}. Notably, the model exhibits both intrinsic and extrinsic HOTI features, where the octupole moment is protected by the \(\mathbf{k}\)-NS symmetries in bulk, and edge polarization is induced either by bulk gap closure affecting SPTPs or edge gap closure affecting BOTPs, depending on boundary terminations. We demonstrate the \( \mathbb{RP}^3 \) HOTI model in a 3D electrical circuit and experimentally observe the octupole corner states by measuring the self-impedance spectra.

\begin{figure}[t]
\centering
\includegraphics[width=\linewidth]{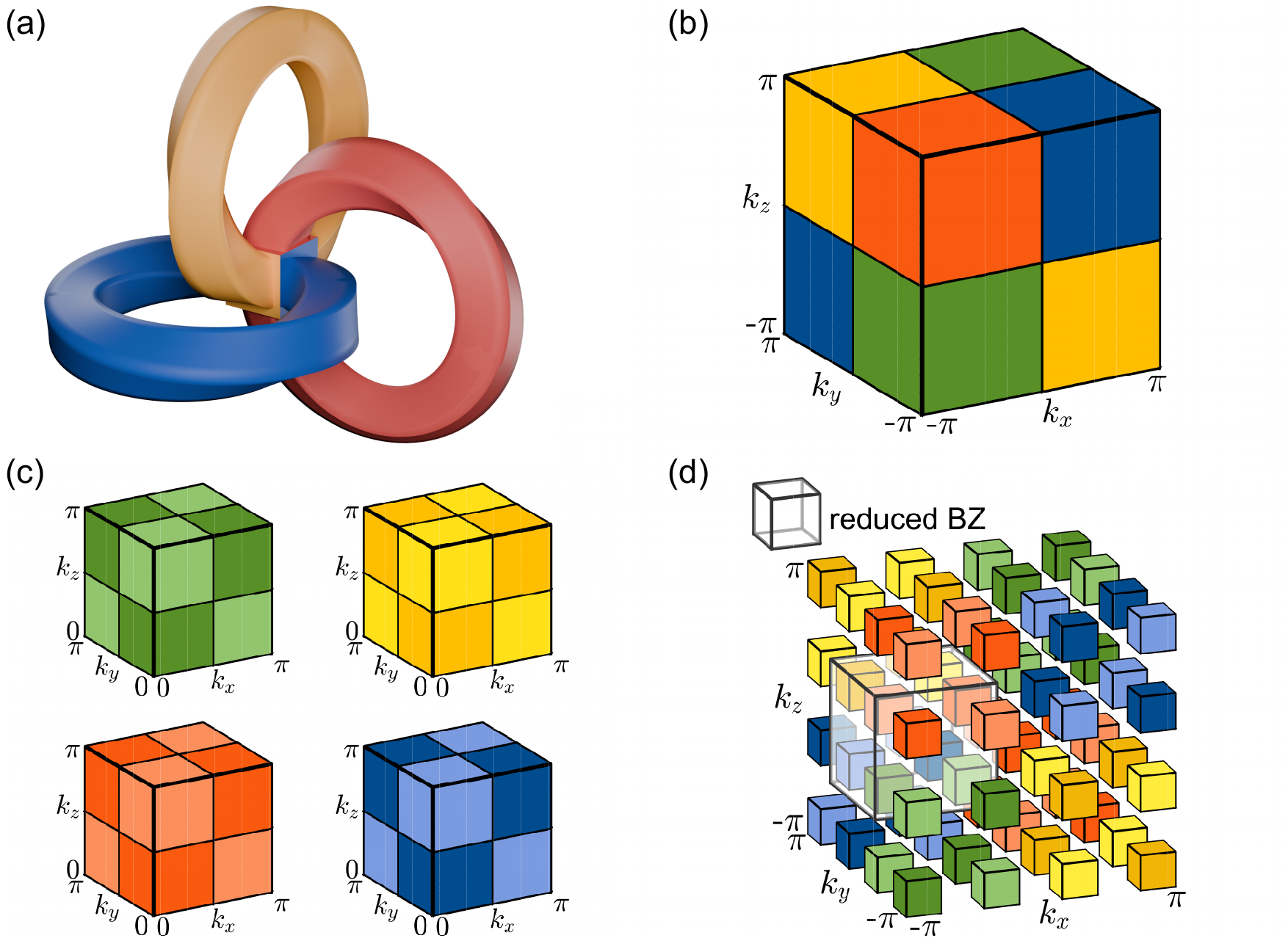}
\caption{Topological construction of the Brillouin \( \mathbb{RP}^3 \) space. (a) Schematic illustration depicting the gluing rules of the Brillouin manifold, with a half-twist operation connecting the opposing faces. (b) The \(\mathbf{k}\)-NS reflection symmetries \( \mathcal{M}_{x} \), \( \mathcal{M}_{y} \), \( \mathcal{M}_{z} \) divide the first BZ into eight segments. (c) The \(\mathbf{k}\)-NS inversion symmetries \( \mathcal{P}_{xy} \), \( \mathcal{P}_{xz} \), \( \mathcal{P}_{yz} \) further divide one segment in (b) into diagonal and off-diagonal pairs. (d) The \(\mathbf{k}\)-NS operators \(\mathcal{M}_{a}\) and \( \mathcal{P}_{ab} \) jointly divide the BZ into 64 blocks, in which a reduced BZ can be defined (semitransparent box) and constructed from any eight uniquely colored blocks.}
\label{fig1}
\end{figure}

\( \mathbb{RP}^3 \) is constructed by adhering the opposing faces of a cube with a half twist [Fig. \ref{fig1}(a)]. Mathematically, it is represented as a unit cube (\([0,1] \times [0,1] \times [0,1]\)) with each pair of opposing faces identified in the specified relation:

\begin{equation} \label{eq1}
\begin{aligned}
(0, y, z) &\sim (1,1-y,1-z), & 0 \leq y,z \leq 1, \\
(x, 0, z) &\sim (1-x,1,1-z), & 0 \leq x,z \leq 1,\\
(x, y, 0) &\sim (1-x,1-y,1), & 0 \leq x,y \leq 1.
\end{aligned}
\end{equation}

Following the designated mapping rule, we realize \( \mathbb{RP}^3 \) in momentum space and derive three \(\mathbf{k}\)-NS reflection operators for the wave vector \( (k_x, k_y, k_z) \) defined as follows:

\begin{equation} \label{eq2}
\begin{aligned}
\mathcal{M}_x : (k_x, k_y, k_z) &\rightarrow (-k_x, k_y + \pi, k_z + \pi), \\
\mathcal{M}_y : (k_x, k_y, k_z) &\rightarrow (k_x + \pi, -k_y, k_z + \pi), \\
\mathcal{M}_z : (k_x, k_y, k_z) &\rightarrow (k_x + \pi, k_y + \pi, -k_z),
\end{aligned}
\end{equation}
where each operator anti-commutes with the others  \cite{RN13,RN14,RN15,PhysRevB.106.L161108}, such that $\{ \mathcal{M}_{a}, \mathcal{M}_{b}\} = 0$ for all $a \neq b$ with $a, b, c \in \{x, y, z\}$. By applying these operators, the first BZ is divided into eight segments, as shown in Fig. \ref{fig1}(b), with diagonally opposing segments forming pairs that are represented with identical colors.  Furthermore, the pairwise combination of these three operators induces novel symmetries, leading to the formulation of \(\mathbf{k}\)-NS inversion operators $\mathcal{P}_{ab} =\mathcal{M}_{a} \mathcal{M}_{b}$:

\begin{equation} \label{eq3}
\begin{aligned}
\mathcal{P}_{xy} : (k_x, k_y, k_z) &\rightarrow (\pi - k_x, \pi - k_y, k_z), \\
\mathcal{P}_{yz} : (k_x, k_y, k_z) &\rightarrow (k_x, \pi - k_y, \pi - k_z), \\
\mathcal{P}_{xz} : (k_x, k_y, k_z) &\rightarrow (\pi - k_x, k_y, \pi - k_z).
\end{aligned}
\end{equation}

A singular operator \( \mathcal{P}_{ab} \) enforces spatial inversion symmetry in the corresponding $a$-$b$ plane within the BZ centered at $(\pm\pi/2, \pm\pi/2)$, resulting in the subdivision of each segment into four blocks along diagonal and off-diagonal pairs [Fig. \ref{fig1}(c)]. The application of the remaining two operators yields analogous subdivisions. Consequently, the \(\mathbf{k}\)-NS symmetric operators \( \mathcal{M} \) and \( \mathcal{P} \) jointly divide the BZ into 64 blocks, where blocks with the same color denote equivalence in the BZ [Fig. \ref{fig1}(d)]. Therefore, a reduced BZ can be defined and constructed from any eight uniquely colored blocks that are enclosed, for instance, by the semitransparent box in Fig. \ref{fig1}(d). The reduced BZ inherits all the topological information from the original BZ, thereby enabling comprehensive analyses of the HOTI, including band structure properties and topological invariants \cite{suppmat}.

\begin{figure}[t]
\centering
\includegraphics[width=\linewidth]{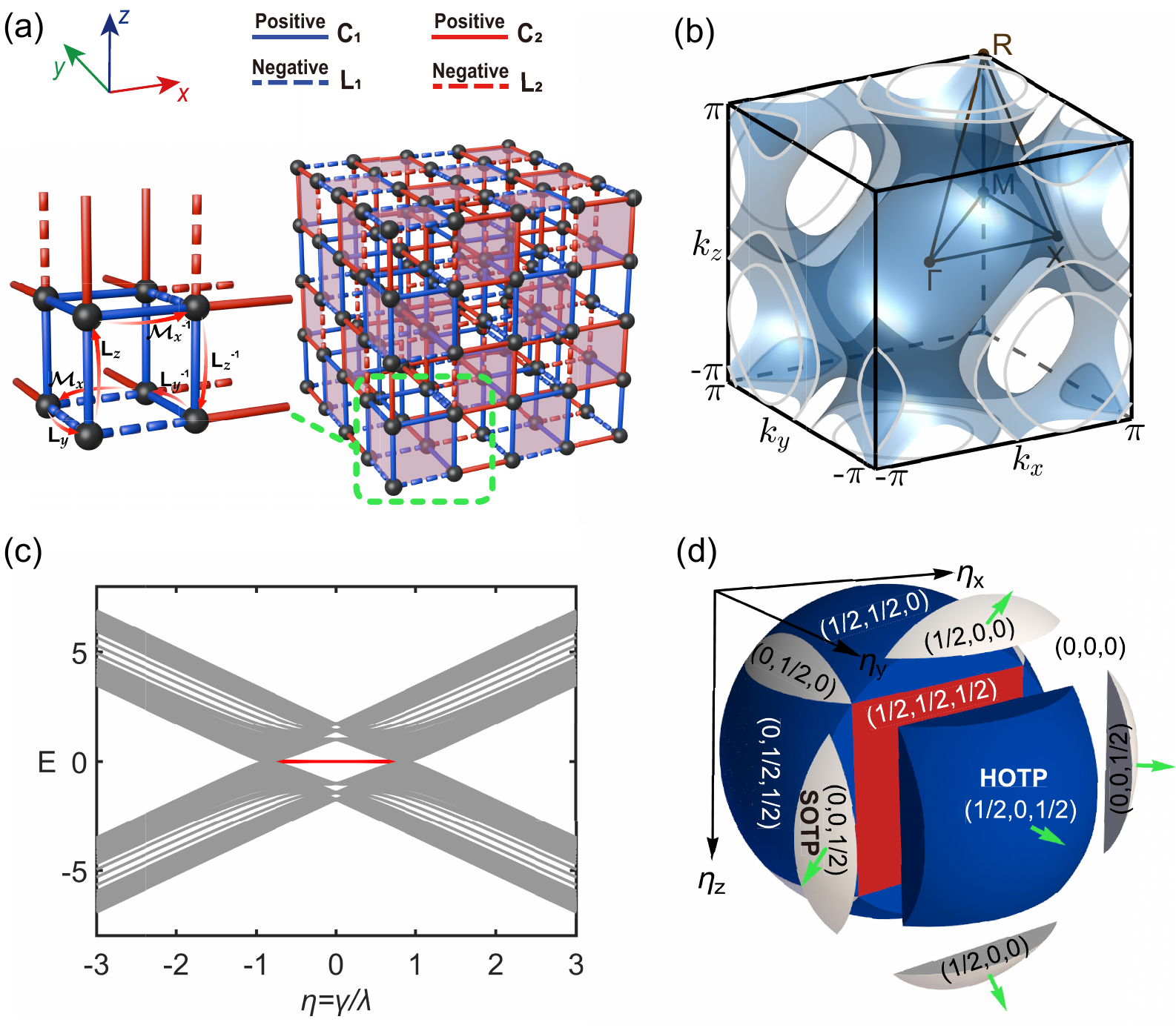}
\caption{Tight-binding model and the topological properties of the HOTI. (a) The lattice with \(2.5\times 2.5\times 2.5\) unit cells, featuring \(\mathbf{k}\)-NS reflection symmetry \(\mathcal{M}_x\). The chessboard \(\pi\)-flux configuration projectively changes the conventional reflection operator \(M_a\) into the \(\mathbf{k}\)-NS one, which anti-commutes with the translation symmetries along the other two directions \(k_b\) and \(k_c\). Solid and dashed lines indicate hoppings with positive and negative signs, respectively; blue lines and red lines represent intra-cell and inter-cell couplings, respectively. (b) Iso-energy contour at E=4 for \(\gamma=1\) and \(\lambda=3.3\). Grey contour lines on the \(k_{b}-k_{c}\) planes are the projections of the blue contour surface at \(k_a=\pm 0.75\pi\). (c) Bulk energy spectrum for a cubic lattice with isotropic coupling strengths and \(N_x=N_y=N_z=10\). Corner states are highlighted by red lines. (d) Phase diagram of the \(\mathbb{RP}^3\) HOTI.} 
\label{fig2}
\end{figure}

To construct the 3D HOTI in Brillouin \( \mathbb{RP}^3 \), we consider a cubic lattice with eight sites as the unit cell, as shown in Fig. \ref{fig2}(a). These eight sites are connected through a specially designed \(\mathbb{Z}_2\) gauge field, with positive (negative) hoppings indicated by solid (dashed) lines. This configuration encloses a \(\pi\)-flux, resulting in an anti-commutative relation between \(\mathbf{k}\)-NS reflection operator \(\mathcal{M}_x\) and translation operators \(\mathbf{L}_y\), \(\mathbf{L}_z\) along the other two directions, respectively. Therefore, in addition to mirror reversion in $k_x$ direction, \(\mathcal{M}_x\) also includes a half-period translation along $k_y$ and $k_z$ simultaneously. Applying these rules to the remaining two directions, we observe in Fig. \ref{fig2}(a) a chessboard \( \pi \)-flux pattern across the $x$-$y$, $x$-$z$, and $y$-$z$ planes. Note that this model significantly differs from the BBH model \cite{RN12, RN3}, in which all plaquettes enclose a \(\pi\)-flux phase.

\begin{figure}[t]
\centering
\includegraphics[width=\linewidth]{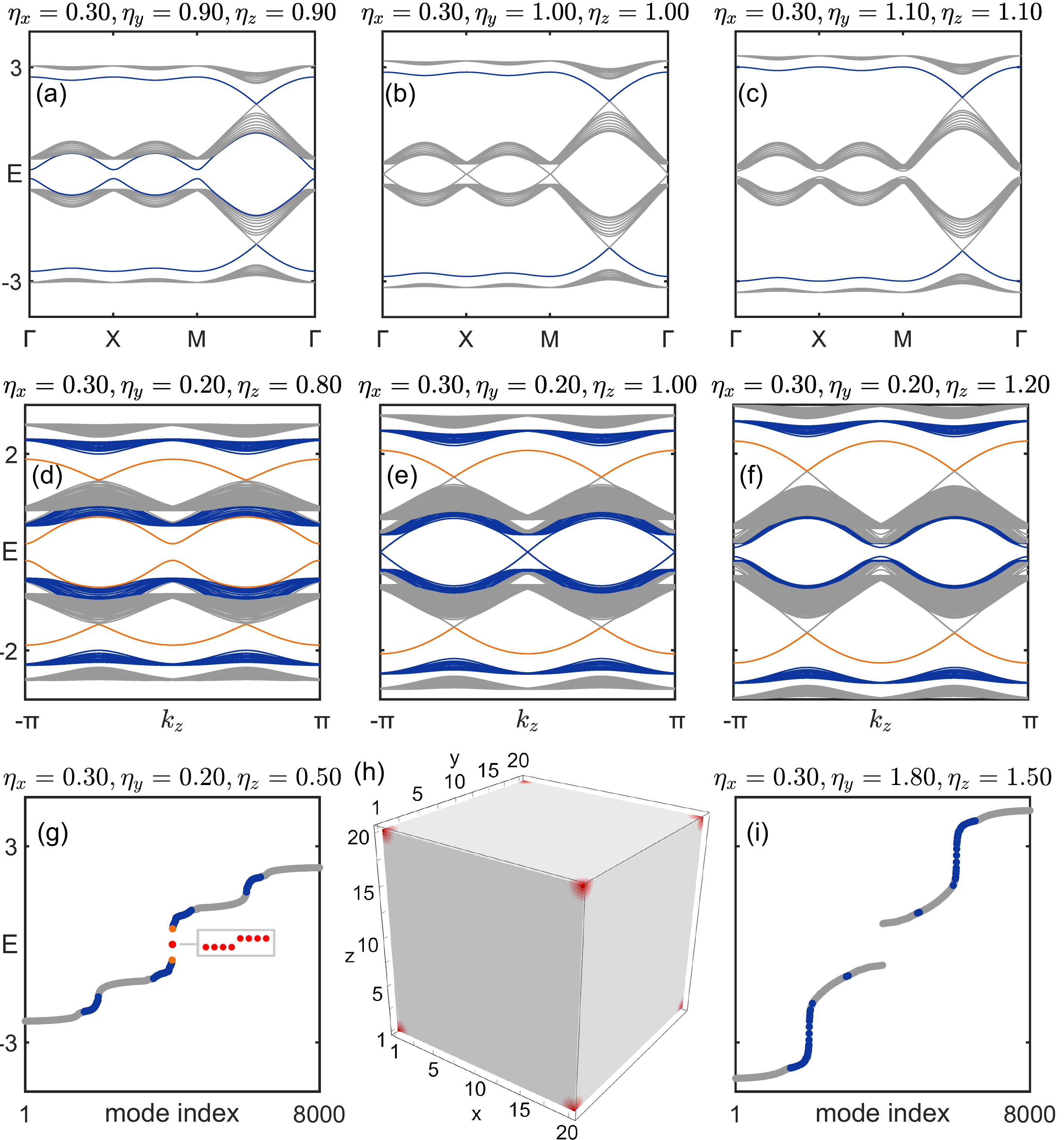}
\caption{ The bulk, surface, and hinge gap closure. (a-c) Phase transition induced by surface gap closure under PBCs in the $y$ and $z$ directions, and OBCs in the $x$ direction. (d-f) Phase transition induced by surface gap closure under PBCs in the $z$ direction and OBCs in the $x$ and $y$ directions. (g-i) Phase transition induced by corner gap closure under full OBCs. The appearance of eight corner modes under full OBCs is shown in (h).}
\label{fig5}
\end{figure}

For convenience, we first assume that the intra-cell coupling strengths \(\gamma\) and the inter-cell coupling strengths \(\lambda\) are isotropic, that is, \(\gamma_{x}=\gamma_{y}=\gamma_{z}=\gamma\) and \(\lambda_{x}=\lambda_{y}=\lambda_{z}=\lambda\). The tight-binding(TB) Hamiltonian can be formulated as follows:
\begin{equation}\label{eq4}
\begin{aligned}
H(k_x, k_y, k_z) =& \lambda (-\cos k_x \Gamma'_3 -\sin k_x \Gamma'_0  + \cos k_y \Gamma'_1-\\&\sin k_y \Gamma'_2  - \cos k_z \Gamma'_0 + \sin k_z \Gamma'_5 ) + \gamma\cdot\zeta,
\end{aligned}
\end{equation}
the matrices \(\Gamma'\) are defined as \(\Gamma'_0 = \sigma_1 \otimes \Gamma_0\), \(\Gamma'_i = \sigma_0 \otimes \Gamma_i \)\((i=1,2,3,4)\), \(\Gamma'_5 = \sigma_2 \otimes \Gamma_0\), where \(\Gamma_0 = \sigma_3\otimes \tau_0\), \(\Gamma_j = \sigma_1 \otimes \tau_j\) \((j=1,2,3)\), \(\Gamma_4 = \sigma_2 \otimes \tau_0\), and $\zeta = \sigma_3 \tau_1  s_0 + \sigma_1  \tau_0  s_0 - \sigma_3  \tau_2 s_2 $, in which \(\sigma\), \(\tau\) and \(s\) are Pauli matrices acting on sites along $x, y, z$ axes respectively. Constrained by the \(\mathbf{k}\)-NS symmetry operators, the band structure in Brillouin \(\mathbb{RP}^3\) displays the corresponding symmetric relations, as evident from the iso-energy contour in Fig. \ref{fig2}(b). This further validates the BZ partition in Fig. \ref{fig1}(d). In addition to the \(\mathbf{k}\)-NS reflection symmetries \( \mathcal{M}_a\) and $\mathcal{P}_{ab}$, \(H(\mathbf{k})\) also retains the conventional inversion symmetry \( \mathcal{I}=\mathcal{M}_x\mathcal{M}_y\mathcal{M}_z\), and the chiral symmetry \(\mathcal{C} {H}(k_x, k_y, k_z) \mathcal{C}^\dagger = -{H}(k_x, k_y, k_z)\). As shown in Fig. \ref{fig2}(c) and throughout Fig. \ref{fig5}, the energy bands appear in pairs at positive and negative energies due to the chiral symmetry \(\mathcal{C}\) of the system \cite{suppmat}.

 Figure \ref{fig2}(c) shows the bulk energy spectrum of the open system as the ratio \(\eta=\gamma/\lambda\) varies. Note that this parameter is isotropic in this case. When \(|\eta| < 1\), in-gap modes emerge at zero energy (red lines), which indicates the presence of octupole corner states. To better understand and characterize the topological properties of the 3D Brillouin \( \mathbb{RP}^3 \) model, a topological invariant of 1/2 can be defined using the nested Wilson loop method \cite{Hu2023HigherOrderTI, RN12}, which suggests a non-trivial topological phase for \( |\eta| < 1 \) and a trivial phase for \( |\eta| > 1 \)  \cite{suppmat}.

Figure \ref{fig2}(d) shows the phase diagram of edge polarization of the 3D \(\mathbb{RP}^3\) HOTI. A sphere in the parameter space (\(\eta_{x}, \eta_{y}, \eta_{z}\)) with a radius of $\sqrt{3}$ is divided into three distinct regions colored in red, gray, and blue, where $\eta_i$ ($i=x, y,z$) represents the ratio of intra-cell to inter-cell hopping strengths along the same direction.
Fixing the inter-cell coupling strengths to $\lambda_x=\lambda_y=\lambda_z=1$ and performing band structure calculations for different $\gamma_x, \gamma_y, \gamma_z$, we find that the bulk bands close when the sum of the squares of the intra-cell hopping strengths equals that of the inter-cell hopping strengths, specifically, when $\gamma_x^2+\gamma_y^2+\gamma_z^2=\lambda_x^2+\lambda_y^2+\lambda_z^2=3$.

We also investigate the impact of varying hopping strengths along periodic directions in semiopen systems, revealing that edge gap closure can induce phase transitions in edge polarization related to BOTPs within the SPTPs. In the first case, with OBCs (open boundary conditions) in the $x$ direction and PBCs(periodic boundary conditions) in the $y$ and $z$ directions, both surface states (blue lines, in the $y-z$ plane) and bulk states (gray lines) coexist, as shown in Fig. \ref{fig5}(a). Simultaneous changes in hopping strengths along the two periodic directions induce phase transitions; specifically, transitioning from the gray region to the red region across the hinge at \(\left|\eta_{y}\right| = \left|\eta_{z}\right| = 1\) results in the closure of the band gap [Fig. \ref{fig5}(b)], and the surface states near zero energy disappear when the gap reopens [Fig. \ref{fig5}(c)]. In the second case, the boundaries in the $x$ and $y$ directions are set as open, and the boundary in the $z$ direction is set as periodic, hinge states (orange lines, in the $z$ direction), surface states (blue lines, in the $x-z$ planes and $y-z$ planes), and bulk states (gray lines)  coexist [Fig. \ref{fig5}(d)] . Varying hopping strengths in the periodic $z-$direction also induces phase transitions; crossing the surface \(\left|\eta_{z}\right|=1\) from the blue region to the red one leads to band gap closure [Fig. \ref{fig5}(e)]. The reopening of the gap is accompanied by the vanishing of the hinge states near zero energy [Fig. \ref{fig5}(f)]. Note that SPTP transition can occur by varying the hopping strength along the open directions in these two semi-open cases \cite{suppmat}. In the third case with the full open system, crossing the point \((\eta_{x},\eta_{y},\eta_{z})=(\pm 1,\,\pm 1,\,\pm 1)\) from the interior to the exterior of the sphere leads to the disappearance of the corner states [Figs. \ref{fig5}(g), (i)].

\begin{figure}[htbp]
\centering
\includegraphics[width=\linewidth]{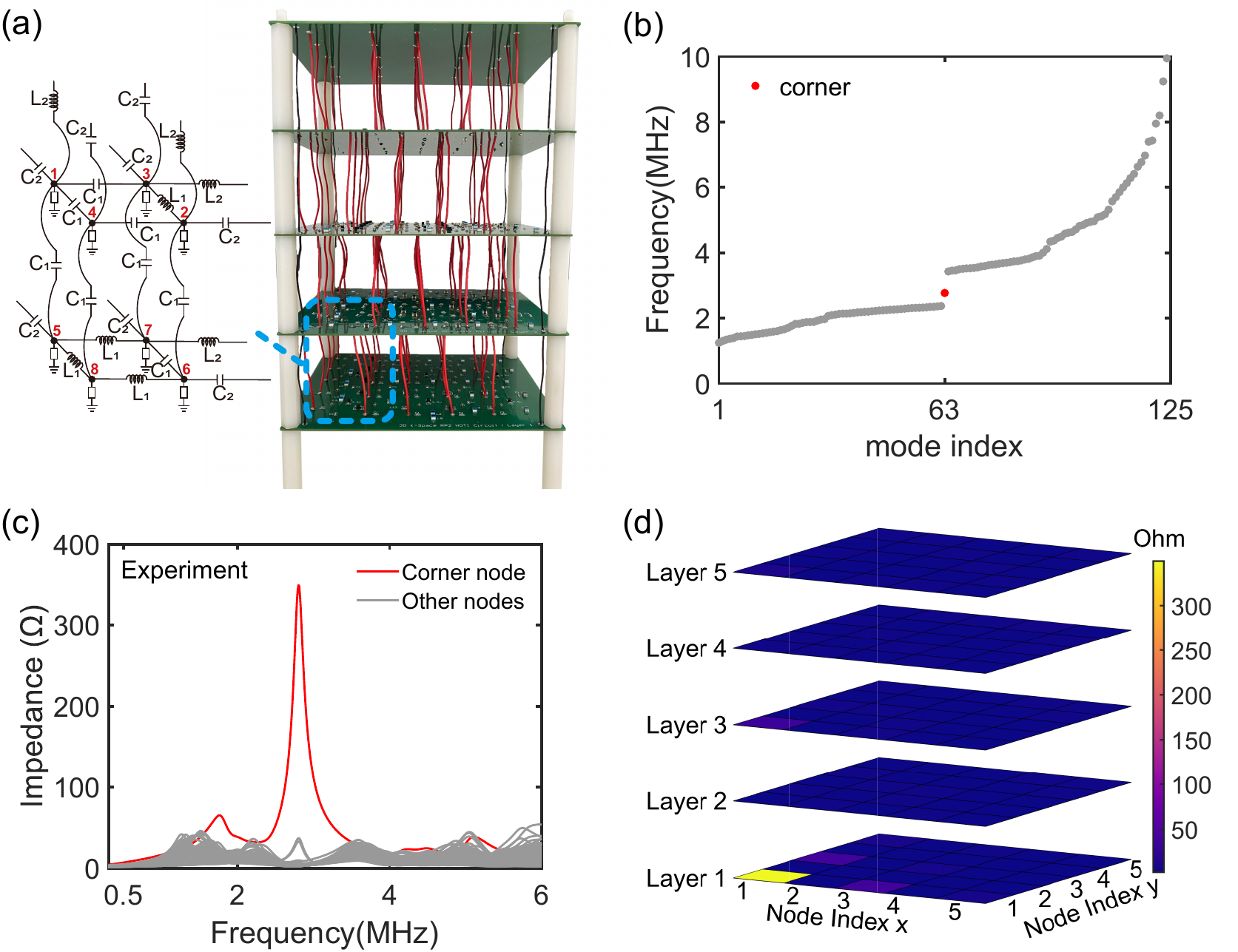}
\caption{Topological properties of the circuit realization. (a) Circuit diagram of the unit cell and the experimental sample. (b) Eigenfrequency of the finite circuit characterized by \({D}\) matrix. The corner mode is highlighted by the red dot. (c) Experimental impedance spectra measured by the vector network analyzer (VNA). (d) Measured impedance distribution at 2.77MHz, clearing demonstrating the presence of the octupole corner state localized at the bottom-left corner. }
\label{fig3}
\end{figure}

The TB model in the quantum electronic system can be directly implemented in the electric circuit by mapping the TB Hamiltonian in Eq. (\ref{eq4}) onto the circuit Laplacian. We realize the \(\mathbb{Z}_2\) gauge connections in circuits by utilizing the opposite phases of the admittance in capacitors and inductors. Two pairs of capacitors and inductors \( (C_1, L_1)\) and \( (C_2=\lambda C_1, L_1=\lambda L_2)\) are employed as the intra-cell and inter-cell couplings in the circuit, respectively, as shown in Fig. \ref{fig3}(a). Note that the boundary circuit nodes should be grounded with additional capacitors and inductors to maintain the same resonant frequency as the bulk nodes, \(\omega_0 = 1 / \sqrt{L_1 C_1} = 1 / \sqrt{L_2 C_2}\)  \cite{suppmat}. In this work, we specify \( C_1=1\text{nF} \), \( C_2=3.3\text{nF}\), \(L_1=3.3\mu\text{H} \), \(L_2=1\mu\text{H}\).

 According to Kirchhoff's current law, we can derive the circuit Laplacian that characterizes the behavior of the circuit as \(\mathbf{J}(\omega) = i\omega \mathbf{C} -i/\omega \mathbf{W}\), where \( \mathbf{C} \) and \(\mathbf{W} \) are the matrices of capacitance and inverse inductance, respectively. Note that, as the admittance of capacitor and inductor cancel at \(\omega_0\), the diagonal terms of \(\mathbf{J}(\omega)\) vanish at \(\omega_0\). Consequently, \(\mathbf{J}(\omega_0)\) takes exactly the form of the Hamiltonian of the quantum electronic system in Eq. (\ref{eq4}), up to a scaling factor of \(i \sqrt{C_1/L_1}\). The eigenfrequencies of the circuit can be obtained through several different approaches, such as using the dynamical matrix \({\mathbf{ D}} = {\mathbf{ C}}^{-\frac{1}{2}}{\mathbf{ W}} {\mathbf{ C}}^{-\frac{1}{2}} \) \cite{RN1, RN9}.

To experimentally demonstrate the octupole corner state induced by the 3D Brillouin  \( \mathbb{RP}^3 \) model, we fabricated a 3D circuit with \( 2.5\times2.5\times2.5 \) unit cells by connecting five layers of printed circuit boards (PCBs) via copper wires. Low DC resistance inductors with a maximum tolerance of 5\% were selected for the experiment to improve the factor of circuit quality while maintaining sufficient precision. Figure \ref{fig3}(b) presents the eigenfrequencies of the finite circuit, where an in-gap mode residing at the resonant frequency \(\omega_0=2.77 \rm{MHz}\), signifies the presence of the octupole corner state. It has been suggested that the eigenstates of the circuit can be accessed by measuring the self-impedance across all circuit nodes at \(\omega_0\), which is proportional to the square of the eigenstates in the quantum electronic system \cite{PhysRevB.110.035106}. In the experiment, we obtained the self-impedance spectra [Fig. \ref{fig3}(c)] independently by measuring the circuit's scattering parameters using a vector network analyzer (Tektronix TTR506A) \cite{suppmat}. Fig. \ref{fig3}(c) shows the measured self-impedance spectra, where a prominent peak at the resonant frequency of 2.77 MHz (red curve) indicates the presence of the topological corner state. Note that due to the configuration with a half-integer number of unit cells in all three dimensions, the current circuit supports only one corner state, with an impedance peak localized at a single corner [Fig. \ref{fig3}(d)]. The results of the 3D \(\mathbb{RP}^3\) topological circuit with an integer number of unit cells are provided in the supplementary material \cite{suppmat}.All experimental results are highly consistent with theoretical calculations \cite{suppmat}. We also verify the phase transitions among the bulk, surface, hinge, and corner states of our 3D HOTI in the circuit system by calculating the band structure and eigenstates of the circuit in both fully open and semi-open scenarios, which aligh with the results from the electronic system shown in Fig. \ref{fig5} \cite{suppmat}.

In conclusion, we experimentally demonstrate a novel octupole topological insulating phase induced by 3D real projective space \(\mathbb{RP}^3\) in momentum space. A \(\pi\)-flux chessboard pattern enforced by the \(\mathbf{k}\)-NS symmetries under the \(\mathbb{Z}_2\) gauge field is shown to give rise to the unconventional features of the 3D \(\mathbb{RP}^3\) HOTI, distinguishing it from the first HOTI with all plaquette treading by \(\pi\)-flux \cite{RN12,RN3}. Note that the current method for implementing the \(\mathbb{Z}_2\) gauge field in electrical circuits with inductors and capacitors allows the direct mapping of the Hamiltonian in the quantum electronic system only at the resonant frequency. This prevents us from measuring the eigenstates of the other modes (e.g., surface and hinge states). Alternative approaches for realizing the negative coupling include the use of negative capacitors with negative impedance convertor (NIC) \cite{chen2003circuits}, or employing a pair of circuit nodes with twist connection \cite{PhysRevB.108.205126}. Both methods enable access to all eigenstates. 


\section*{Acknowledgements}
The authors thank Prof. Shaojie Ma for fruitful discussions and Ruiwen Shao for experimental support. This work acknowledges funding from the National Key Research and Development Program of China under Grant No.\ 2022YFA1404903, 2023YFB3811504 and the National Natural Science Foundation of China under Grant No.\  U22A2001, 62201136, 62288101, and 61905101, the start-up Research Fund of Southeast University No.\ RF1028623117.


\begin{thebibliography}{58}%
\makeatletter
\providecommand \@ifxundefined [1]{%
 \@ifx{#1\undefined}
}%
\providecommand \@ifnum [1]{%
 \ifnum #1\expandafter \@firstoftwo
 \else \expandafter \@secondoftwo
 \fi
}%
\providecommand \@ifx [1]{%
 \ifx #1\expandafter \@firstoftwo
 \else \expandafter \@secondoftwo
 \fi
}%
\providecommand \natexlab [1]{#1}%
\providecommand \enquote  [1]{``#1''}%
\providecommand \bibnamefont  [1]{#1}%
\providecommand \bibfnamefont [1]{#1}%
\providecommand \citenamefont [1]{#1}%
\providecommand \href@noop [0]{\@secondoftwo}%
\providecommand \href [0]{\begingroup \@sanitize@url \@href}%
\providecommand \@href[1]{\@@startlink{#1}\@@href}%
\providecommand \@@href[1]{\endgroup#1\@@endlink}%
\providecommand \@sanitize@url [0]{\catcode `\\12\catcode `\$12\catcode `\&12\catcode `\#12\catcode `\^12\catcode `\_12\catcode `\%12\relax}%
\providecommand \@@startlink[1]{}%
\providecommand \@@endlink[0]{}%
\providecommand \url  [0]{\begingroup\@sanitize@url \@url }%
\providecommand \@url [1]{\endgroup\@href {#1}{\urlprefix }}%
\providecommand \urlprefix  [0]{URL }%
\providecommand \Eprint [0]{\href }%
\providecommand \doibase [0]{http://dx.doi.org/}%
\providecommand \selectlanguage [0]{\@gobble}%
\providecommand \bibinfo  [0]{\@secondoftwo}%
\providecommand \bibfield  [0]{\@secondoftwo}%
\providecommand \translation [1]{[#1]}%
\providecommand \BibitemOpen [0]{}%
\providecommand \bibitemStop [0]{}%
\providecommand \bibitemNoStop [0]{.\EOS\space}%
\providecommand \EOS [0]{\spacefactor3000\relax}%
\providecommand \BibitemShut  [1]{\csname bibitem#1\endcsname}%
\let\auto@bib@innerbib\@empty
\bibitem [{\citenamefont {Franca}\ \emph {et~al.}(2018)\citenamefont {Franca}, \citenamefont {van~den Brink},\ and\ \citenamefont {Fulga}}]{PhysRevB.98.201114}%
  \BibitemOpen
  \bibfield  {author} {\bibinfo {author} {\bibfnamefont {S.}~\bibnamefont {Franca}}, \bibinfo {author} {\bibfnamefont {J.}~\bibnamefont {van~den Brink}}, \ and\ \bibinfo {author} {\bibfnamefont {I.~C.}\ \bibnamefont {Fulga}},\ }\href {\doibase 10.1103/PhysRevB.98.201114} {\bibfield  {journal} {\bibinfo  {journal} {Phys. Rev. B}\ }\textbf {\bibinfo {volume} {98}},\ \bibinfo {pages} {201114} (\bibinfo {year} {2018})}\BibitemShut {NoStop}%
\bibitem [{\citenamefont {Xie}\ \emph {et~al.}(2018)\citenamefont {Xie}, \citenamefont {Wang}, \citenamefont {Wang}, \citenamefont {Zhu}, \citenamefont {Jiang}, \citenamefont {Lu},\ and\ \citenamefont {Chen}}]{PhysRevB.98.205147}%
  \BibitemOpen
  \bibfield  {author} {\bibinfo {author} {\bibfnamefont {B.-Y.}\ \bibnamefont {Xie}}, \bibinfo {author} {\bibfnamefont {H.-F.}\ \bibnamefont {Wang}}, \bibinfo {author} {\bibfnamefont {H.-X.}\ \bibnamefont {Wang}}, \bibinfo {author} {\bibfnamefont {X.-Y.}\ \bibnamefont {Zhu}}, \bibinfo {author} {\bibfnamefont {J.-H.}\ \bibnamefont {Jiang}}, \bibinfo {author} {\bibfnamefont {M.-H.}\ \bibnamefont {Lu}}, \ and\ \bibinfo {author} {\bibfnamefont {Y.-F.}\ \bibnamefont {Chen}},\ }\href {\doibase 10.1103/PhysRevB.98.205147} {\bibfield  {journal} {\bibinfo  {journal} {Phys. Rev. B}\ }\textbf {\bibinfo {volume} {98}},\ \bibinfo {pages} {205147} (\bibinfo {year} {2018})}\BibitemShut {NoStop}%
\bibitem [{\citenamefont {Lumer}\ \emph {et~al.}(2013)\citenamefont {Lumer}, \citenamefont {Plotnik}, \citenamefont {Rechtsman},\ and\ \citenamefont {Segev}}]{PhysRevLett.111.243905}%
  \BibitemOpen
  \bibfield  {author} {\bibinfo {author} {\bibfnamefont {Y.}~\bibnamefont {Lumer}}, \bibinfo {author} {\bibfnamefont {Y.}~\bibnamefont {Plotnik}}, \bibinfo {author} {\bibfnamefont {M.~C.}\ \bibnamefont {Rechtsman}}, \ and\ \bibinfo {author} {\bibfnamefont {M.}~\bibnamefont {Segev}},\ }\href {\doibase 10.1103/PhysRevLett.111.243905} {\bibfield  {journal} {\bibinfo  {journal} {Phys. Rev. Lett.}\ }\textbf {\bibinfo {volume} {111}},\ \bibinfo {pages} {243905} (\bibinfo {year} {2013})}\BibitemShut {NoStop}%
\bibitem [{\citenamefont {Gao}\ \emph {et~al.}(2015)\citenamefont {Gao}, \citenamefont {Lawrence}, \citenamefont {Yang}, \citenamefont {Liu}, \citenamefont {Fang}, \citenamefont {B\'eri}, \citenamefont {Li},\ and\ \citenamefont {Zhang}}]{PhysRevLett.114.037402}%
  \BibitemOpen
  \bibfield  {author} {\bibinfo {author} {\bibfnamefont {W.}~\bibnamefont {Gao}}, \bibinfo {author} {\bibfnamefont {M.}~\bibnamefont {Lawrence}}, \bibinfo {author} {\bibfnamefont {B.}~\bibnamefont {Yang}}, \bibinfo {author} {\bibfnamefont {F.}~\bibnamefont {Liu}}, \bibinfo {author} {\bibfnamefont {F.}~\bibnamefont {Fang}}, \bibinfo {author} {\bibfnamefont {B.}~\bibnamefont {B\'eri}}, \bibinfo {author} {\bibfnamefont {J.}~\bibnamefont {Li}}, \ and\ \bibinfo {author} {\bibfnamefont {S.}~\bibnamefont {Zhang}},\ }\href {\doibase 10.1103/PhysRevLett.114.037402} {\bibfield  {journal} {\bibinfo  {journal} {Phys. Rev. Lett.}\ }\textbf {\bibinfo {volume} {114}},\ \bibinfo {pages} {037402} (\bibinfo {year} {2015})}\BibitemShut {NoStop}%
\bibitem [{\citenamefont {Noh}\ \emph {et~al.}(2018)\citenamefont {Noh}, \citenamefont {Huang}, \citenamefont {Chen},\ and\ \citenamefont {Rechtsman}}]{PhysRevLett.120.063902}%
  \BibitemOpen
  \bibfield  {author} {\bibinfo {author} {\bibfnamefont {J.}~\bibnamefont {Noh}}, \bibinfo {author} {\bibfnamefont {S.}~\bibnamefont {Huang}}, \bibinfo {author} {\bibfnamefont {K.~P.}\ \bibnamefont {Chen}}, \ and\ \bibinfo {author} {\bibfnamefont {M.~C.}\ \bibnamefont {Rechtsman}},\ }\href {\doibase 10.1103/PhysRevLett.120.063902} {\bibfield  {journal} {\bibinfo  {journal} {Phys. Rev. Lett.}\ }\textbf {\bibinfo {volume} {120}},\ \bibinfo {pages} {063902} (\bibinfo {year} {2018})}\BibitemShut {NoStop}%
\bibitem [{\citenamefont {Yang}\ \emph {et~al.}(2015)\citenamefont {Yang}, \citenamefont {Gao}, \citenamefont {Shi}, \citenamefont {Lin}, \citenamefont {Gao}, \citenamefont {Chong},\ and\ \citenamefont {Zhang}}]{PhysRevLett.114.114301}%
  \BibitemOpen
  \bibfield  {author} {\bibinfo {author} {\bibfnamefont {Z.}~\bibnamefont {Yang}}, \bibinfo {author} {\bibfnamefont {F.}~\bibnamefont {Gao}}, \bibinfo {author} {\bibfnamefont {X.}~\bibnamefont {Shi}}, \bibinfo {author} {\bibfnamefont {X.}~\bibnamefont {Lin}}, \bibinfo {author} {\bibfnamefont {Z.}~\bibnamefont {Gao}}, \bibinfo {author} {\bibfnamefont {Y.}~\bibnamefont {Chong}}, \ and\ \bibinfo {author} {\bibfnamefont {B.}~\bibnamefont {Zhang}},\ }\href {\doibase 10.1103/PhysRevLett.114.114301} {\bibfield  {journal} {\bibinfo  {journal} {Phys. Rev. Lett.}\ }\textbf {\bibinfo {volume} {114}},\ \bibinfo {pages} {114301} (\bibinfo {year} {2015})}\BibitemShut {NoStop}%
\bibitem [{\citenamefont {Xia}\ \emph {et~al.}(2017)\citenamefont {Xia}, \citenamefont {Liu}, \citenamefont {Huang}, \citenamefont {Dai}, \citenamefont {Jiao}, \citenamefont {Zang}, \citenamefont {Yu}, \citenamefont {Zheng},\ and\ \citenamefont {Liu}}]{PhysRevB.96.094106}%
  \BibitemOpen
  \bibfield  {author} {\bibinfo {author} {\bibfnamefont {B.-Z.}\ \bibnamefont {Xia}}, \bibinfo {author} {\bibfnamefont {T.-T.}\ \bibnamefont {Liu}}, \bibinfo {author} {\bibfnamefont {G.-L.}\ \bibnamefont {Huang}}, \bibinfo {author} {\bibfnamefont {H.-Q.}\ \bibnamefont {Dai}}, \bibinfo {author} {\bibfnamefont {J.-R.}\ \bibnamefont {Jiao}}, \bibinfo {author} {\bibfnamefont {X.-G.}\ \bibnamefont {Zang}}, \bibinfo {author} {\bibfnamefont {D.-J.}\ \bibnamefont {Yu}}, \bibinfo {author} {\bibfnamefont {S.-J.}\ \bibnamefont {Zheng}}, \ and\ \bibinfo {author} {\bibfnamefont {J.}~\bibnamefont {Liu}},\ }\href {\doibase 10.1103/PhysRevB.96.094106} {\bibfield  {journal} {\bibinfo  {journal} {Phys. Rev. B}\ }\textbf {\bibinfo {volume} {96}},\ \bibinfo {pages} {094106} (\bibinfo {year} {2017})}\BibitemShut {NoStop}%
\bibitem [{\citenamefont {Deng}\ \emph {et~al.}(2017)\citenamefont {Deng}, \citenamefont {Ge}, \citenamefont {Tian}, \citenamefont {Lu},\ and\ \citenamefont {Jing}}]{PhysRevB.96.184305}%
  \BibitemOpen
  \bibfield  {author} {\bibinfo {author} {\bibfnamefont {Y.}~\bibnamefont {Deng}}, \bibinfo {author} {\bibfnamefont {H.}~\bibnamefont {Ge}}, \bibinfo {author} {\bibfnamefont {Y.}~\bibnamefont {Tian}}, \bibinfo {author} {\bibfnamefont {M.}~\bibnamefont {Lu}}, \ and\ \bibinfo {author} {\bibfnamefont {Y.}~\bibnamefont {Jing}},\ }\href {\doibase 10.1103/PhysRevB.96.184305} {\bibfield  {journal} {\bibinfo  {journal} {Phys. Rev. B}\ }\textbf {\bibinfo {volume} {96}},\ \bibinfo {pages} {184305} (\bibinfo {year} {2017})}\BibitemShut {NoStop}%
\bibitem [{\citenamefont {Xue}\ \emph {et~al.}(2019)\citenamefont {Xue}, \citenamefont {Yang}, \citenamefont {Liu}, \citenamefont {Gao}, \citenamefont {Chong},\ and\ \citenamefont {Zhang}}]{PhysRevLett.122.244301}%
  \BibitemOpen
  \bibfield  {author} {\bibinfo {author} {\bibfnamefont {H.}~\bibnamefont {Xue}}, \bibinfo {author} {\bibfnamefont {Y.}~\bibnamefont {Yang}}, \bibinfo {author} {\bibfnamefont {G.}~\bibnamefont {Liu}}, \bibinfo {author} {\bibfnamefont {F.}~\bibnamefont {Gao}}, \bibinfo {author} {\bibfnamefont {Y.}~\bibnamefont {Chong}}, \ and\ \bibinfo {author} {\bibfnamefont {B.}~\bibnamefont {Zhang}},\ }\href {\doibase 10.1103/PhysRevLett.122.244301} {\bibfield  {journal} {\bibinfo  {journal} {Phys. Rev. Lett.}\ }\textbf {\bibinfo {volume} {122}},\ \bibinfo {pages} {244301} (\bibinfo {year} {2019})}\BibitemShut {NoStop}%
\bibitem [{\citenamefont {Qi}\ \emph {et~al.}(2020)\citenamefont {Qi}, \citenamefont {Qiu}, \citenamefont {Xiao}, \citenamefont {He}, \citenamefont {Ke},\ and\ \citenamefont {Liu}}]{PhysRevLett.124.206601}%
  \BibitemOpen
  \bibfield  {author} {\bibinfo {author} {\bibfnamefont {Y.}~\bibnamefont {Qi}}, \bibinfo {author} {\bibfnamefont {C.}~\bibnamefont {Qiu}}, \bibinfo {author} {\bibfnamefont {M.}~\bibnamefont {Xiao}}, \bibinfo {author} {\bibfnamefont {H.}~\bibnamefont {He}}, \bibinfo {author} {\bibfnamefont {M.}~\bibnamefont {Ke}}, \ and\ \bibinfo {author} {\bibfnamefont {Z.}~\bibnamefont {Liu}},\ }\href {\doibase 10.1103/PhysRevLett.124.206601} {\bibfield  {journal} {\bibinfo  {journal} {Phys. Rev. Lett.}\ }\textbf {\bibinfo {volume} {124}},\ \bibinfo {pages} {206601} (\bibinfo {year} {2020})}\BibitemShut {NoStop}%
\bibitem [{\citenamefont {Rocklin}\ \emph {et~al.}(2016)\citenamefont {Rocklin}, \citenamefont {Chen}, \citenamefont {Falk}, \citenamefont {Vitelli},\ and\ \citenamefont {Lubensky}}]{PhysRevLett.116.135503}%
  \BibitemOpen
  \bibfield  {author} {\bibinfo {author} {\bibfnamefont {D.~Z.}\ \bibnamefont {Rocklin}}, \bibinfo {author} {\bibfnamefont {B.~G.-g.}\ \bibnamefont {Chen}}, \bibinfo {author} {\bibfnamefont {M.}~\bibnamefont {Falk}}, \bibinfo {author} {\bibfnamefont {V.}~\bibnamefont {Vitelli}}, \ and\ \bibinfo {author} {\bibfnamefont {T.~C.}\ \bibnamefont {Lubensky}},\ }\href {\doibase 10.1103/PhysRevLett.116.135503} {\bibfield  {journal} {\bibinfo  {journal} {Phys. Rev. Lett.}\ }\textbf {\bibinfo {volume} {116}},\ \bibinfo {pages} {135503} (\bibinfo {year} {2016})}\BibitemShut {NoStop}%
\bibitem [{\citenamefont {Chen}\ \emph {et~al.}(2019)\citenamefont {Chen}, \citenamefont {Yao}, \citenamefont {Nassar},\ and\ \citenamefont {Huang}}]{PhysRevApplied.11.044029}%
  \BibitemOpen
  \bibfield  {author} {\bibinfo {author} {\bibfnamefont {H.}~\bibnamefont {Chen}}, \bibinfo {author} {\bibfnamefont {L.}~\bibnamefont {Yao}}, \bibinfo {author} {\bibfnamefont {H.}~\bibnamefont {Nassar}}, \ and\ \bibinfo {author} {\bibfnamefont {G.}~\bibnamefont {Huang}},\ }\href {\doibase 10.1103/PhysRevApplied.11.044029} {\bibfield  {journal} {\bibinfo  {journal} {Phys. Rev. Appl.}\ }\textbf {\bibinfo {volume} {11}},\ \bibinfo {pages} {044029} (\bibinfo {year} {2019})}\BibitemShut {NoStop}%
\bibitem [{\citenamefont {Ezawa}(2018)}]{PhysRevB.98.201402}%
  \BibitemOpen
  \bibfield  {author} {\bibinfo {author} {\bibfnamefont {M.}~\bibnamefont {Ezawa}},\ }\href {\doibase 10.1103/PhysRevB.98.201402} {\bibfield  {journal} {\bibinfo  {journal} {Phys. Rev. B}\ }\textbf {\bibinfo {volume} {98}},\ \bibinfo {pages} {201402} (\bibinfo {year} {2018})}\BibitemShut {NoStop}%
\bibitem [{\citenamefont {Imhof}\ \emph {et~al.}(2018)\citenamefont {Imhof}, \citenamefont {Berger}, \citenamefont {Bayer}, \citenamefont {Brehm}, \citenamefont {Molenkamp}, \citenamefont {Kiessling}, \citenamefont {Schindler}, \citenamefont {Lee}, \citenamefont {Greiter}, \citenamefont {Neupert},\ and\ \citenamefont {Thomale}}]{RN1}%
  \BibitemOpen
  \bibfield  {author} {\bibinfo {author} {\bibfnamefont {S.}~\bibnamefont {Imhof}}, \bibinfo {author} {\bibfnamefont {C.}~\bibnamefont {Berger}}, \bibinfo {author} {\bibfnamefont {F.}~\bibnamefont {Bayer}}, \bibinfo {author} {\bibfnamefont {J.}~\bibnamefont {Brehm}}, \bibinfo {author} {\bibfnamefont {L.~W.}\ \bibnamefont {Molenkamp}}, \bibinfo {author} {\bibfnamefont {T.}~\bibnamefont {Kiessling}}, \bibinfo {author} {\bibfnamefont {F.}~\bibnamefont {Schindler}}, \bibinfo {author} {\bibfnamefont {C.~H.}\ \bibnamefont {Lee}}, \bibinfo {author} {\bibfnamefont {M.}~\bibnamefont {Greiter}}, \bibinfo {author} {\bibfnamefont {T.}~\bibnamefont {Neupert}}, \ and\ \bibinfo {author} {\bibfnamefont {R.}~\bibnamefont {Thomale}},\ }\href {\doibase 10.1038/s41567-018-0246-1} {\bibfield  {journal} {\bibinfo  {journal} {Nat. Phys.}\ }\textbf {\bibinfo {volume} {14}},\ \bibinfo {pages} {925} (\bibinfo {year} {2018})}\BibitemShut {NoStop}%
\bibitem [{\citenamefont {Ezawa}(2019)}]{PhysRevB.99.121411}%
  \BibitemOpen
  \bibfield  {author} {\bibinfo {author} {\bibfnamefont {M.}~\bibnamefont {Ezawa}},\ }\href {\doibase 10.1103/PhysRevB.99.121411} {\bibfield  {journal} {\bibinfo  {journal} {Phys. Rev. B}\ }\textbf {\bibinfo {volume} {99}},\ \bibinfo {pages} {121411} (\bibinfo {year} {2019})}\BibitemShut {NoStop}%
\bibitem [{\citenamefont {Hofmann}\ \emph {et~al.}(2020)\citenamefont {Hofmann}, \citenamefont {Helbig}, \citenamefont {Schindler}, \citenamefont {Salgo}, \citenamefont {Brzezi\ifmmode~\acute{n}\else \'{n}\fi{}ska}, \citenamefont {Greiter}, \citenamefont {Kiessling}, \citenamefont {Wolf}, \citenamefont {Vollhardt}, \citenamefont {Kaba\ifmmode~\check{s}\else \v{s}\fi{}i}, \citenamefont {Lee}, \citenamefont {Bilu\ifmmode \check{s}\else \v{s}\fi{}i\ifmmode~\acute{c}\else \'{c}\fi{}}, \citenamefont {Thomale},\ and\ \citenamefont {Neupert}}]{PhysRevResearch.2.023265}%
  \BibitemOpen
  \bibfield  {author} {\bibinfo {author} {\bibfnamefont {T.}~\bibnamefont {Hofmann}}, \bibinfo {author} {\bibfnamefont {T.}~\bibnamefont {Helbig}}, \bibinfo {author} {\bibfnamefont {F.}~\bibnamefont {Schindler}}, \bibinfo {author} {\bibfnamefont {N.}~\bibnamefont {Salgo}}, \bibinfo {author} {\bibfnamefont {M.}~\bibnamefont {Brzezi\ifmmode~\acute{n}\else \'{n}\fi{}ska}}, \bibinfo {author} {\bibfnamefont {M.}~\bibnamefont {Greiter}}, \bibinfo {author} {\bibfnamefont {T.}~\bibnamefont {Kiessling}}, \bibinfo {author} {\bibfnamefont {D.}~\bibnamefont {Wolf}}, \bibinfo {author} {\bibfnamefont {A.}~\bibnamefont {Vollhardt}}, \bibinfo {author} {\bibfnamefont {A.}~\bibnamefont {Kaba\ifmmode~\check{s}\else \v{s}\fi{}i}}, \bibinfo {author} {\bibfnamefont {C.~H.}\ \bibnamefont {Lee}}, \bibinfo {author} {\bibfnamefont {A.}~\bibnamefont {Bilu\ifmmode \check{s}\else \v{s}\fi{}i\ifmmode~\acute{c}\else \'{c}\fi{}}}, \bibinfo {author} {\bibfnamefont {R.}~\bibnamefont {Thomale}}, \ and\ \bibinfo {author} {\bibfnamefont
  {T.}~\bibnamefont {Neupert}},\ }\href {\doibase 10.1103/PhysRevResearch.2.023265} {\bibfield  {journal} {\bibinfo  {journal} {Phys. Rev. Res.}\ }\textbf {\bibinfo {volume} {2}},\ \bibinfo {pages} {023265} (\bibinfo {year} {2020})}\BibitemShut {NoStop}%
\bibitem [{\citenamefont {Li}\ \emph {et~al.}(2020)\citenamefont {Li}, \citenamefont {Fu}, \citenamefont {Hu}, \citenamefont {Li},\ and\ \citenamefont {Shen}}]{PhysRevLett.125.166801}%
  \BibitemOpen
  \bibfield  {author} {\bibinfo {author} {\bibfnamefont {C.-A.}\ \bibnamefont {Li}}, \bibinfo {author} {\bibfnamefont {B.}~\bibnamefont {Fu}}, \bibinfo {author} {\bibfnamefont {Z.-A.}\ \bibnamefont {Hu}}, \bibinfo {author} {\bibfnamefont {J.}~\bibnamefont {Li}}, \ and\ \bibinfo {author} {\bibfnamefont {S.-Q.}\ \bibnamefont {Shen}},\ }\href {\doibase 10.1103/PhysRevLett.125.166801} {\bibfield  {journal} {\bibinfo  {journal} {Phys. Rev. Lett.}\ }\textbf {\bibinfo {volume} {125}},\ \bibinfo {pages} {166801} (\bibinfo {year} {2020})}\BibitemShut {NoStop}%
\bibitem [{\citenamefont {Zhang}\ \emph {et~al.}(2021)\citenamefont {Zhang}, \citenamefont {Zou}, \citenamefont {Pei}, \citenamefont {He}, \citenamefont {Bao}, \citenamefont {Sun},\ and\ \citenamefont {Zhang}}]{PhysRevLett.126.146802}%
  \BibitemOpen
  \bibfield  {author} {\bibinfo {author} {\bibfnamefont {W.}~\bibnamefont {Zhang}}, \bibinfo {author} {\bibfnamefont {D.}~\bibnamefont {Zou}}, \bibinfo {author} {\bibfnamefont {Q.}~\bibnamefont {Pei}}, \bibinfo {author} {\bibfnamefont {W.}~\bibnamefont {He}}, \bibinfo {author} {\bibfnamefont {J.}~\bibnamefont {Bao}}, \bibinfo {author} {\bibfnamefont {H.}~\bibnamefont {Sun}}, \ and\ \bibinfo {author} {\bibfnamefont {X.}~\bibnamefont {Zhang}},\ }\href {\doibase 10.1103/PhysRevLett.126.146802} {\bibfield  {journal} {\bibinfo  {journal} {Phys. Rev. Lett.}\ }\textbf {\bibinfo {volume} {126}},\ \bibinfo {pages} {146802} (\bibinfo {year} {2021})}\BibitemShut {NoStop}%
\bibitem [{\citenamefont {Wang}\ \emph {et~al.}(2020{\natexlab{a}})\citenamefont {Wang}, \citenamefont {Price}, \citenamefont {Zhang},\ and\ \citenamefont {Chong}}]{RN2}%
  \BibitemOpen
  \bibfield  {author} {\bibinfo {author} {\bibfnamefont {Y.}~\bibnamefont {Wang}}, \bibinfo {author} {\bibfnamefont {H.~M.}\ \bibnamefont {Price}}, \bibinfo {author} {\bibfnamefont {B.}~\bibnamefont {Zhang}}, \ and\ \bibinfo {author} {\bibfnamefont {Y.~D.}\ \bibnamefont {Chong}},\ }\href {\doibase 10.1038/s41467-020-15940-3} {\bibfield  {journal} {\bibinfo  {journal} {Nat. Commun.}\ }\textbf {\bibinfo {volume} {11}},\ \bibinfo {pages} {2356} (\bibinfo {year} {2020}{\natexlab{a}})}\BibitemShut {NoStop}%
\bibitem [{\citenamefont {Liu}\ \emph {et~al.}(2021)\citenamefont {Liu}, \citenamefont {Ma}, \citenamefont {Shao}, \citenamefont {Zhang}, \citenamefont {Yang}, \citenamefont {Navarro-Cía}, \citenamefont {Cui},\ and\ \citenamefont {Zhang}}]{Liu_2021}%
  \BibitemOpen
  \bibfield  {author} {\bibinfo {author} {\bibfnamefont {S.}~\bibnamefont {Liu}}, \bibinfo {author} {\bibfnamefont {S.}~\bibnamefont {Ma}}, \bibinfo {author} {\bibfnamefont {R.}~\bibnamefont {Shao}}, \bibinfo {author} {\bibfnamefont {L.}~\bibnamefont {Zhang}}, \bibinfo {author} {\bibfnamefont {B.}~\bibnamefont {Yang}}, \bibinfo {author} {\bibfnamefont {M.}~\bibnamefont {Navarro-Cía}}, \bibinfo {author} {\bibfnamefont {T.~J.}\ \bibnamefont {Cui}}, \ and\ \bibinfo {author} {\bibfnamefont {S.}~\bibnamefont {Zhang}},\ }\href {\doibase 10.1088/1367-2630/ac2755} {\bibfield  {journal} {\bibinfo  {journal} {New J. Phys.}\ }\textbf {\bibinfo {volume} {23}},\ \bibinfo {pages} {103005} (\bibinfo {year} {2021})}\BibitemShut {NoStop}%
\bibitem [{\citenamefont {Liu}\ \emph {et~al.}(2020{\natexlab{a}})\citenamefont {Liu}, \citenamefont {Ma}, \citenamefont {Yang}, \citenamefont {Zhang}, \citenamefont {Gao}, \citenamefont {Xiang}, \citenamefont {Cui},\ and\ \citenamefont {Zhang}}]{PhysRevApplied.13.014047}%
  \BibitemOpen
  \bibfield  {author} {\bibinfo {author} {\bibfnamefont {S.}~\bibnamefont {Liu}}, \bibinfo {author} {\bibfnamefont {S.}~\bibnamefont {Ma}}, \bibinfo {author} {\bibfnamefont {C.}~\bibnamefont {Yang}}, \bibinfo {author} {\bibfnamefont {L.}~\bibnamefont {Zhang}}, \bibinfo {author} {\bibfnamefont {W.}~\bibnamefont {Gao}}, \bibinfo {author} {\bibfnamefont {Y.~J.}\ \bibnamefont {Xiang}}, \bibinfo {author} {\bibfnamefont {T.~J.}\ \bibnamefont {Cui}}, \ and\ \bibinfo {author} {\bibfnamefont {S.}~\bibnamefont {Zhang}},\ }\href {\doibase 10.1103/PhysRevApplied.13.014047} {\bibfield  {journal} {\bibinfo  {journal} {Phys. Rev. Appl.}\ }\textbf {\bibinfo {volume} {13}},\ \bibinfo {pages} {014047} (\bibinfo {year} {2020}{\natexlab{a}})}\BibitemShut {NoStop}%
\bibitem [{\citenamefont {Chen}\ \emph {et~al.}(2020)\citenamefont {Chen}, \citenamefont {Chen}, \citenamefont {Gao}, \citenamefont {Zhou},\ and\ \citenamefont {Xu}}]{PhysRevLett.124.036803}%
  \BibitemOpen
  \bibfield  {author} {\bibinfo {author} {\bibfnamefont {R.}~\bibnamefont {Chen}}, \bibinfo {author} {\bibfnamefont {C.-Z.}\ \bibnamefont {Chen}}, \bibinfo {author} {\bibfnamefont {J.-H.}\ \bibnamefont {Gao}}, \bibinfo {author} {\bibfnamefont {B.}~\bibnamefont {Zhou}}, \ and\ \bibinfo {author} {\bibfnamefont {D.-H.}\ \bibnamefont {Xu}},\ }\href {\doibase 10.1103/PhysRevLett.124.036803} {\bibfield  {journal} {\bibinfo  {journal} {Phys. Rev. Lett.}\ }\textbf {\bibinfo {volume} {124}},\ \bibinfo {pages} {036803} (\bibinfo {year} {2020})}\BibitemShut {NoStop}%
\bibitem [{\citenamefont {Fan}\ \emph {et~al.}(2019)\citenamefont {Fan}, \citenamefont {Xia}, \citenamefont {Tong}, \citenamefont {Zheng},\ and\ \citenamefont {Yu}}]{PhysRevLett.122.204301}%
  \BibitemOpen
  \bibfield  {author} {\bibinfo {author} {\bibfnamefont {H.}~\bibnamefont {Fan}}, \bibinfo {author} {\bibfnamefont {B.}~\bibnamefont {Xia}}, \bibinfo {author} {\bibfnamefont {L.}~\bibnamefont {Tong}}, \bibinfo {author} {\bibfnamefont {S.}~\bibnamefont {Zheng}}, \ and\ \bibinfo {author} {\bibfnamefont {D.}~\bibnamefont {Yu}},\ }\href {\doibase 10.1103/PhysRevLett.122.204301} {\bibfield  {journal} {\bibinfo  {journal} {Phys. Rev. Lett.}\ }\textbf {\bibinfo {volume} {122}},\ \bibinfo {pages} {204301} (\bibinfo {year} {2019})}\BibitemShut {NoStop}%
\bibitem [{\citenamefont {Liu}\ \emph {et~al.}(2020{\natexlab{b}})\citenamefont {Liu}, \citenamefont {Ma}, \citenamefont {Zhang}, \citenamefont {Zhang}, \citenamefont {Yang}, \citenamefont {You}, \citenamefont {Gao}, \citenamefont {Xiang}, \citenamefont {Cui},\ and\ \citenamefont {Zhang}}]{RN3}%
  \BibitemOpen
  \bibfield  {author} {\bibinfo {author} {\bibfnamefont {S.}~\bibnamefont {Liu}}, \bibinfo {author} {\bibfnamefont {S.}~\bibnamefont {Ma}}, \bibinfo {author} {\bibfnamefont {Q.}~\bibnamefont {Zhang}}, \bibinfo {author} {\bibfnamefont {L.}~\bibnamefont {Zhang}}, \bibinfo {author} {\bibfnamefont {C.}~\bibnamefont {Yang}}, \bibinfo {author} {\bibfnamefont {O.}~\bibnamefont {You}}, \bibinfo {author} {\bibfnamefont {W.}~\bibnamefont {Gao}}, \bibinfo {author} {\bibfnamefont {Y.}~\bibnamefont {Xiang}}, \bibinfo {author} {\bibfnamefont {T.~J.}\ \bibnamefont {Cui}}, \ and\ \bibinfo {author} {\bibfnamefont {S.}~\bibnamefont {Zhang}},\ }\href {\doibase 10.1038/s41377-020-00381-w} {\bibfield  {journal} {\bibinfo  {journal} {Light: Sci. Appl.}\ }\textbf {\bibinfo {volume} {9}},\ \bibinfo {pages} {145} (\bibinfo {year} {2020}{\natexlab{b}})}\BibitemShut {NoStop}%
\bibitem [{\citenamefont {Peterson}\ \emph {et~al.}(2018)\citenamefont {Peterson}, \citenamefont {Benalcazar}, \citenamefont {Hughes},\ and\ \citenamefont {Bahl}}]{RN16}%
  \BibitemOpen
  \bibfield  {author} {\bibinfo {author} {\bibfnamefont {C.~W.}\ \bibnamefont {Peterson}}, \bibinfo {author} {\bibfnamefont {W.~A.}\ \bibnamefont {Benalcazar}}, \bibinfo {author} {\bibfnamefont {T.~L.}\ \bibnamefont {Hughes}}, \ and\ \bibinfo {author} {\bibfnamefont {G.}~\bibnamefont {Bahl}},\ }\href {\doibase 10.1038/nature25777} {\bibfield  {journal} {\bibinfo  {journal} {Nature}\ }\textbf {\bibinfo {volume} {555}},\ \bibinfo {pages} {346} (\bibinfo {year} {2018})}\BibitemShut {NoStop}%
\bibitem [{\citenamefont {Peterson}\ \emph {et~al.}(2021)\citenamefont {Peterson}, \citenamefont {Li}, \citenamefont {Jiang}, \citenamefont {Hughes},\ and\ \citenamefont {Bahl}}]{RN17}%
  \BibitemOpen
  \bibfield  {author} {\bibinfo {author} {\bibfnamefont {C.~W.}\ \bibnamefont {Peterson}}, \bibinfo {author} {\bibfnamefont {T.}~\bibnamefont {Li}}, \bibinfo {author} {\bibfnamefont {W.}~\bibnamefont {Jiang}}, \bibinfo {author} {\bibfnamefont {T.~L.}\ \bibnamefont {Hughes}}, \ and\ \bibinfo {author} {\bibfnamefont {G.}~\bibnamefont {Bahl}},\ }\href {\doibase 10.1038/s41586-020-03117-3} {\bibfield  {journal} {\bibinfo  {journal} {Nature}\ }\textbf {\bibinfo {volume} {589}},\ \bibinfo {pages} {376} (\bibinfo {year} {2021})}\BibitemShut {NoStop}%
\bibitem [{\citenamefont {Zhang}\ \emph {et~al.}(2020{\natexlab{a}})\citenamefont {Zhang}, \citenamefont {Wu},\ and\ \citenamefont {Das~Sarma}}]{PhysRevLett.124.136407}%
  \BibitemOpen
  \bibfield  {author} {\bibinfo {author} {\bibfnamefont {R.-X.}\ \bibnamefont {Zhang}}, \bibinfo {author} {\bibfnamefont {F.}~\bibnamefont {Wu}}, \ and\ \bibinfo {author} {\bibfnamefont {S.}~\bibnamefont {Das~Sarma}},\ }\href {\doibase 10.1103/PhysRevLett.124.136407} {\bibfield  {journal} {\bibinfo  {journal} {Phys. Rev. Lett.}\ }\textbf {\bibinfo {volume} {124}},\ \bibinfo {pages} {136407} (\bibinfo {year} {2020}{\natexlab{a}})}\BibitemShut {NoStop}%
\bibitem [{\citenamefont {Wang}\ \emph {et~al.}(2020{\natexlab{b}})\citenamefont {Wang}, \citenamefont {Lin}, \citenamefont {Jiang}, \citenamefont {Guo},\ and\ \citenamefont {Jiang}}]{PhysRevLett.125.146401}%
  \BibitemOpen
  \bibfield  {author} {\bibinfo {author} {\bibfnamefont {H.-X.}\ \bibnamefont {Wang}}, \bibinfo {author} {\bibfnamefont {Z.-K.}\ \bibnamefont {Lin}}, \bibinfo {author} {\bibfnamefont {B.}~\bibnamefont {Jiang}}, \bibinfo {author} {\bibfnamefont {G.-Y.}\ \bibnamefont {Guo}}, \ and\ \bibinfo {author} {\bibfnamefont {J.-H.}\ \bibnamefont {Jiang}},\ }\href {\doibase 10.1103/PhysRevLett.125.146401} {\bibfield  {journal} {\bibinfo  {journal} {Phys. Rev. Lett.}\ }\textbf {\bibinfo {volume} {125}},\ \bibinfo {pages} {146401} (\bibinfo {year} {2020}{\natexlab{b}})}\BibitemShut {NoStop}%
\bibitem [{\citenamefont {Xu}\ \emph {et~al.}(2019)\citenamefont {Xu}, \citenamefont {Song}, \citenamefont {Wang}, \citenamefont {Weng},\ and\ \citenamefont {Dai}}]{PhysRevLett.122.256402}%
  \BibitemOpen
  \bibfield  {author} {\bibinfo {author} {\bibfnamefont {Y.}~\bibnamefont {Xu}}, \bibinfo {author} {\bibfnamefont {Z.}~\bibnamefont {Song}}, \bibinfo {author} {\bibfnamefont {Z.}~\bibnamefont {Wang}}, \bibinfo {author} {\bibfnamefont {H.}~\bibnamefont {Weng}}, \ and\ \bibinfo {author} {\bibfnamefont {X.}~\bibnamefont {Dai}},\ }\href {\doibase 10.1103/PhysRevLett.122.256402} {\bibfield  {journal} {\bibinfo  {journal} {Phys. Rev. Lett.}\ }\textbf {\bibinfo {volume} {122}},\ \bibinfo {pages} {256402} (\bibinfo {year} {2019})}\BibitemShut {NoStop}%
\bibitem [{\citenamefont {Shiozaki}\ and\ \citenamefont {Sato}(2014)}]{PhysRevB.90.165114}%
  \BibitemOpen
  \bibfield  {author} {\bibinfo {author} {\bibfnamefont {K.}~\bibnamefont {Shiozaki}}\ and\ \bibinfo {author} {\bibfnamefont {M.}~\bibnamefont {Sato}},\ }\href {\doibase 10.1103/PhysRevB.90.165114} {\bibfield  {journal} {\bibinfo  {journal} {Phys. Rev. B}\ }\textbf {\bibinfo {volume} {90}},\ \bibinfo {pages} {165114} (\bibinfo {year} {2014})}\BibitemShut {NoStop}%
\bibitem [{\citenamefont {Chiu}\ \emph {et~al.}(2016)\citenamefont {Chiu}, \citenamefont {Teo}, \citenamefont {Schnyder},\ and\ \citenamefont {Ryu}}]{RevModPhys.88.035005}%
  \BibitemOpen
  \bibfield  {author} {\bibinfo {author} {\bibfnamefont {C.-K.}\ \bibnamefont {Chiu}}, \bibinfo {author} {\bibfnamefont {J.~C.~Y.}\ \bibnamefont {Teo}}, \bibinfo {author} {\bibfnamefont {A.~P.}\ \bibnamefont {Schnyder}}, \ and\ \bibinfo {author} {\bibfnamefont {S.}~\bibnamefont {Ryu}},\ }\href {\doibase 10.1103/RevModPhys.88.035005} {\bibfield  {journal} {\bibinfo  {journal} {Rev. Mod. Phys.}\ }\textbf {\bibinfo {volume} {88}},\ \bibinfo {pages} {035005} (\bibinfo {year} {2016})}\BibitemShut {NoStop}%
\bibitem [{\citenamefont {Chiu}\ and\ \citenamefont {Schnyder}(2014)}]{PhysRevB.90.205136}%
  \BibitemOpen
  \bibfield  {author} {\bibinfo {author} {\bibfnamefont {C.-K.}\ \bibnamefont {Chiu}}\ and\ \bibinfo {author} {\bibfnamefont {A.~P.}\ \bibnamefont {Schnyder}},\ }\href {\doibase 10.1103/PhysRevB.90.205136} {\bibfield  {journal} {\bibinfo  {journal} {Phys. Rev. B}\ }\textbf {\bibinfo {volume} {90}},\ \bibinfo {pages} {205136} (\bibinfo {year} {2014})}\BibitemShut {NoStop}%
\bibitem [{\citenamefont {Kawabata}\ \emph {et~al.}(2019)\citenamefont {Kawabata}, \citenamefont {Shiozaki}, \citenamefont {Ueda},\ and\ \citenamefont {Sato}}]{PhysRevX.9.041015}%
  \BibitemOpen
  \bibfield  {author} {\bibinfo {author} {\bibfnamefont {K.}~\bibnamefont {Kawabata}}, \bibinfo {author} {\bibfnamefont {K.}~\bibnamefont {Shiozaki}}, \bibinfo {author} {\bibfnamefont {M.}~\bibnamefont {Ueda}}, \ and\ \bibinfo {author} {\bibfnamefont {M.}~\bibnamefont {Sato}},\ }\href {\doibase 10.1103/PhysRevX.9.041015} {\bibfield  {journal} {\bibinfo  {journal} {Phys. Rev. X}\ }\textbf {\bibinfo {volume} {9}},\ \bibinfo {pages} {041015} (\bibinfo {year} {2019})}\BibitemShut {NoStop}%
\bibitem [{\citenamefont {Po}\ \emph {et~al.}(2018)\citenamefont {Po}, \citenamefont {Watanabe},\ and\ \citenamefont {Vishwanath}}]{PhysRevLett.121.126402}%
  \BibitemOpen
  \bibfield  {author} {\bibinfo {author} {\bibfnamefont {H.~C.}\ \bibnamefont {Po}}, \bibinfo {author} {\bibfnamefont {H.}~\bibnamefont {Watanabe}}, \ and\ \bibinfo {author} {\bibfnamefont {A.}~\bibnamefont {Vishwanath}},\ }\href {\doibase 10.1103/PhysRevLett.121.126402} {\bibfield  {journal} {\bibinfo  {journal} {Phys. Rev. Lett.}\ }\textbf {\bibinfo {volume} {121}},\ \bibinfo {pages} {126402} (\bibinfo {year} {2018})}\BibitemShut {NoStop}%
\bibitem [{\citenamefont {Zohar}\ \emph {et~al.}(2017)\citenamefont {Zohar}, \citenamefont {Farace}, \citenamefont {Reznik},\ and\ \citenamefont {Cirac}}]{PhysRevLett.118.070501}%
  \BibitemOpen
  \bibfield  {author} {\bibinfo {author} {\bibfnamefont {E.}~\bibnamefont {Zohar}}, \bibinfo {author} {\bibfnamefont {A.}~\bibnamefont {Farace}}, \bibinfo {author} {\bibfnamefont {B.}~\bibnamefont {Reznik}}, \ and\ \bibinfo {author} {\bibfnamefont {J.~I.}\ \bibnamefont {Cirac}},\ }\href {\doibase 10.1103/PhysRevLett.118.070501} {\bibfield  {journal} {\bibinfo  {journal} {Phys. Rev. Lett.}\ }\textbf {\bibinfo {volume} {118}},\ \bibinfo {pages} {070501} (\bibinfo {year} {2017})}\BibitemShut {NoStop}%
\bibitem [{\citenamefont {Feng}\ \emph {et~al.}(2007)\citenamefont {Feng}, \citenamefont {Zhang},\ and\ \citenamefont {Xiang}}]{PhysRevLett.98.087204}%
  \BibitemOpen
  \bibfield  {author} {\bibinfo {author} {\bibfnamefont {X.-Y.}\ \bibnamefont {Feng}}, \bibinfo {author} {\bibfnamefont {G.-M.}\ \bibnamefont {Zhang}}, \ and\ \bibinfo {author} {\bibfnamefont {T.}~\bibnamefont {Xiang}},\ }\href {\doibase 10.1103/PhysRevLett.98.087204} {\bibfield  {journal} {\bibinfo  {journal} {Phys. Rev. Lett.}\ }\textbf {\bibinfo {volume} {98}},\ \bibinfo {pages} {087204} (\bibinfo {year} {2007})}\BibitemShut {NoStop}%
\bibitem [{\citenamefont {Borla}\ \emph {et~al.}(2020)\citenamefont {Borla}, \citenamefont {Verresen}, \citenamefont {Grusdt},\ and\ \citenamefont {Moroz}}]{PhysRevLett.124.120503}%
  \BibitemOpen
  \bibfield  {author} {\bibinfo {author} {\bibfnamefont {U.}~\bibnamefont {Borla}}, \bibinfo {author} {\bibfnamefont {R.}~\bibnamefont {Verresen}}, \bibinfo {author} {\bibfnamefont {F.}~\bibnamefont {Grusdt}}, \ and\ \bibinfo {author} {\bibfnamefont {S.}~\bibnamefont {Moroz}},\ }\href {\doibase 10.1103/PhysRevLett.124.120503} {\bibfield  {journal} {\bibinfo  {journal} {Phys. Rev. Lett.}\ }\textbf {\bibinfo {volume} {124}},\ \bibinfo {pages} {120503} (\bibinfo {year} {2020})}\BibitemShut {NoStop}%
\bibitem [{\citenamefont {Yao}\ and\ \citenamefont {Qi}(2010)}]{PhysRevLett.105.080501}%
  \BibitemOpen
  \bibfield  {author} {\bibinfo {author} {\bibfnamefont {H.}~\bibnamefont {Yao}}\ and\ \bibinfo {author} {\bibfnamefont {X.-L.}\ \bibnamefont {Qi}},\ }\href {\doibase 10.1103/PhysRevLett.105.080501} {\bibfield  {journal} {\bibinfo  {journal} {Phys. Rev. Lett.}\ }\textbf {\bibinfo {volume} {105}},\ \bibinfo {pages} {080501} (\bibinfo {year} {2010})}\BibitemShut {NoStop}%
\bibitem [{\citenamefont {Xue}\ \emph {et~al.}(2022)\citenamefont {Xue}, \citenamefont {Wang}, \citenamefont {Huang}, \citenamefont {Cheng}, \citenamefont {Yu}, \citenamefont {Foo}, \citenamefont {Zhao}, \citenamefont {Yang},\ and\ \citenamefont {Zhang}}]{PhysRevLett.128.116802}%
  \BibitemOpen
  \bibfield  {author} {\bibinfo {author} {\bibfnamefont {H.}~\bibnamefont {Xue}}, \bibinfo {author} {\bibfnamefont {Z.}~\bibnamefont {Wang}}, \bibinfo {author} {\bibfnamefont {Y.-X.}\ \bibnamefont {Huang}}, \bibinfo {author} {\bibfnamefont {Z.}~\bibnamefont {Cheng}}, \bibinfo {author} {\bibfnamefont {L.}~\bibnamefont {Yu}}, \bibinfo {author} {\bibfnamefont {Y.~X.}\ \bibnamefont {Foo}}, \bibinfo {author} {\bibfnamefont {Y.~X.}\ \bibnamefont {Zhao}}, \bibinfo {author} {\bibfnamefont {S.~A.}\ \bibnamefont {Yang}}, \ and\ \bibinfo {author} {\bibfnamefont {B.}~\bibnamefont {Zhang}},\ }\href {\doibase 10.1103/PhysRevLett.128.116802} {\bibfield  {journal} {\bibinfo  {journal} {Phys. Rev. Lett.}\ }\textbf {\bibinfo {volume} {128}},\ \bibinfo {pages} {116802} (\bibinfo {year} {2022})}\BibitemShut {NoStop}%
\bibitem [{\citenamefont {Cui}\ \emph {et~al.}(2022)\citenamefont {Cui}, \citenamefont {Zhang}, \citenamefont {Zhang},\ and\ \citenamefont {Chan}}]{PhysRevLett.129.043902}%
  \BibitemOpen
  \bibfield  {author} {\bibinfo {author} {\bibfnamefont {X.}~\bibnamefont {Cui}}, \bibinfo {author} {\bibfnamefont {R.-Y.}\ \bibnamefont {Zhang}}, \bibinfo {author} {\bibfnamefont {Z.-Q.}\ \bibnamefont {Zhang}}, \ and\ \bibinfo {author} {\bibfnamefont {C.~T.}\ \bibnamefont {Chan}},\ }\href {\doibase 10.1103/PhysRevLett.129.043902} {\bibfield  {journal} {\bibinfo  {journal} {Phys. Rev. Lett.}\ }\textbf {\bibinfo {volume} {129}},\ \bibinfo {pages} {043902} (\bibinfo {year} {2022})}\BibitemShut {NoStop}%
\bibitem [{\citenamefont {Chen}\ \emph {et~al.}(2022)\citenamefont {Chen}, \citenamefont {Yang},\ and\ \citenamefont {Zhao}}]{RN8}%
  \BibitemOpen
  \bibfield  {author} {\bibinfo {author} {\bibfnamefont {Z.~Y.}\ \bibnamefont {Chen}}, \bibinfo {author} {\bibfnamefont {S.~A.}\ \bibnamefont {Yang}}, \ and\ \bibinfo {author} {\bibfnamefont {Y.~X.}\ \bibnamefont {Zhao}},\ }\href {\doibase 10.1038/s41467-022-29953-7} {\bibfield  {journal} {\bibinfo  {journal} {Nat. Commun.}\ }\textbf {\bibinfo {volume} {13}},\ \bibinfo {pages} {2215} (\bibinfo {year} {2022})}\BibitemShut {NoStop}%
\bibitem [{\citenamefont {Li}\ \emph {et~al.}(2023)\citenamefont {Li}, \citenamefont {Sun}, \citenamefont {Zhang}, \citenamefont {Guo},\ and\ \citenamefont {Trauzettel}}]{PhysRevB.108.235412}%
  \BibitemOpen
  \bibfield  {author} {\bibinfo {author} {\bibfnamefont {C.-A.}\ \bibnamefont {Li}}, \bibinfo {author} {\bibfnamefont {J.}~\bibnamefont {Sun}}, \bibinfo {author} {\bibfnamefont {S.-B.}\ \bibnamefont {Zhang}}, \bibinfo {author} {\bibfnamefont {H.}~\bibnamefont {Guo}}, \ and\ \bibinfo {author} {\bibfnamefont {B.}~\bibnamefont {Trauzettel}},\ }\href {\doibase 10.1103/PhysRevB.108.235412} {\bibfield  {journal} {\bibinfo  {journal} {Phys. Rev. B}\ }\textbf {\bibinfo {volume} {108}},\ \bibinfo {pages} {235412} (\bibinfo {year} {2023})}\BibitemShut {NoStop}%
\bibitem [{\citenamefont {Li}\ \emph {et~al.}(2022)\citenamefont {Li}, \citenamefont {Du}, \citenamefont {Zhang}, \citenamefont {Li}, \citenamefont {Fan}, \citenamefont {Zhang},\ and\ \citenamefont {Qiu}}]{PhysRevLett.128.116803}%
  \BibitemOpen
  \bibfield  {author} {\bibinfo {author} {\bibfnamefont {T.}~\bibnamefont {Li}}, \bibinfo {author} {\bibfnamefont {J.}~\bibnamefont {Du}}, \bibinfo {author} {\bibfnamefont {Q.}~\bibnamefont {Zhang}}, \bibinfo {author} {\bibfnamefont {Y.}~\bibnamefont {Li}}, \bibinfo {author} {\bibfnamefont {X.}~\bibnamefont {Fan}}, \bibinfo {author} {\bibfnamefont {F.}~\bibnamefont {Zhang}}, \ and\ \bibinfo {author} {\bibfnamefont {C.}~\bibnamefont {Qiu}},\ }\href {\doibase 10.1103/PhysRevLett.128.116803} {\bibfield  {journal} {\bibinfo  {journal} {Phys. Rev. Lett.}\ }\textbf {\bibinfo {volume} {128}},\ \bibinfo {pages} {116803} (\bibinfo {year} {2022})}\BibitemShut {NoStop}%
\bibitem [{\citenamefont {Hu}\ \emph {et~al.}(2023)\citenamefont {Hu}, \citenamefont {Zhuang},\ and\ \citenamefont {Yang}}]{Hu2023HigherOrderTI}%
  \BibitemOpen
  \bibfield  {author} {\bibinfo {author} {\bibfnamefont {J.}~\bibnamefont {Hu}}, \bibinfo {author} {\bibfnamefont {S.}~\bibnamefont {Zhuang}}, \ and\ \bibinfo {author} {\bibfnamefont {Y.}~\bibnamefont {Yang}},\ }\href {https://api.semanticscholar.org/CorpusID:259262209} {\bibfield  {journal} {\bibinfo  {journal} {Phys. Rev. Lett.}\ }\textbf {\bibinfo {volume} {132 21}},\ \bibinfo {pages} {213801} (\bibinfo {year} {2023})}\BibitemShut {NoStop}%
\bibitem [{\citenamefont {Shang}\ \emph {et~al.}(2024)\citenamefont {Shang}, \citenamefont {Liu}, \citenamefont {Jiang}, \citenamefont {Shao}, \citenamefont {Zang}, \citenamefont {Lee}, \citenamefont {Thomale}, \citenamefont {Manchon}, \citenamefont {Cui},\ and\ \citenamefont {Schwingenschlögl}}]{RN5}%
  \BibitemOpen
  \bibfield  {author} {\bibinfo {author} {\bibfnamefont {C.}~\bibnamefont {Shang}}, \bibinfo {author} {\bibfnamefont {S.}~\bibnamefont {Liu}}, \bibinfo {author} {\bibfnamefont {C.}~\bibnamefont {Jiang}}, \bibinfo {author} {\bibfnamefont {R.}~\bibnamefont {Shao}}, \bibinfo {author} {\bibfnamefont {X.}~\bibnamefont {Zang}}, \bibinfo {author} {\bibfnamefont {C.~H.}\ \bibnamefont {Lee}}, \bibinfo {author} {\bibfnamefont {R.}~\bibnamefont {Thomale}}, \bibinfo {author} {\bibfnamefont {A.}~\bibnamefont {Manchon}}, \bibinfo {author} {\bibfnamefont {T.~J.}\ \bibnamefont {Cui}}, \ and\ \bibinfo {author} {\bibfnamefont {U.}~\bibnamefont {Schwingenschlögl}},\ }\href {\doibase https://doi.org/10.1002/advs.202303222} {\bibfield  {journal} {\bibinfo  {journal} {Adv. Sci,}\ }\textbf {\bibinfo {volume} {11}},\ \bibinfo {pages} {2303222} (\bibinfo {year} {2024})}\BibitemShut {NoStop}%
\bibitem [{\citenamefont {Zhenxiao}\ \emph {et~al.}(2024)\citenamefont {Zhenxiao}, \citenamefont {Linyun}, \citenamefont {Jien}, \citenamefont {Yan}, \citenamefont {Xiang}, \citenamefont {Bei}, \citenamefont {Jingming}, \citenamefont {Jiuyang}, \citenamefont {Xueqin}, \citenamefont {Weiyin}, \citenamefont {Ce}, \citenamefont {Perry~Ping}, \citenamefont {Yihao}, \citenamefont {Hongsheng}, \citenamefont {Kexin}, \citenamefont {Gui-Geng}, \citenamefont {Zhengyou},\ and\ \citenamefont {Zhen}}]{RN35}%
  \BibitemOpen
  \bibfield  {author} {\bibinfo {author} {\bibfnamefont {Z.}~\bibnamefont {Zhenxiao}}, \bibinfo {author} {\bibfnamefont {Y.}~\bibnamefont {Linyun}}, \bibinfo {author} {\bibfnamefont {W.}~\bibnamefont {Jien}}, \bibinfo {author} {\bibfnamefont {M.}~\bibnamefont {Yan}}, \bibinfo {author} {\bibfnamefont {X.}~\bibnamefont {Xiang}}, \bibinfo {author} {\bibfnamefont {Y.}~\bibnamefont {Bei}}, \bibinfo {author} {\bibfnamefont {C.}~\bibnamefont {Jingming}}, \bibinfo {author} {\bibfnamefont {L.}~\bibnamefont {Jiuyang}}, \bibinfo {author} {\bibfnamefont {H.}~\bibnamefont {Xueqin}}, \bibinfo {author} {\bibfnamefont {D.}~\bibnamefont {Weiyin}}, \bibinfo {author} {\bibfnamefont {S.}~\bibnamefont {Ce}}, \bibinfo {author} {\bibfnamefont {S.}~\bibnamefont {Perry~Ping}}, \bibinfo {author} {\bibfnamefont {Y.}~\bibnamefont {Yihao}}, \bibinfo {author} {\bibfnamefont {C.}~\bibnamefont {Hongsheng}}, \bibinfo {author} {\bibfnamefont {X.}~\bibnamefont {Kexin}}, \bibinfo {author} {\bibfnamefont {L.}~\bibnamefont {Gui-Geng}}, \bibinfo
  {author} {\bibfnamefont {L.}~\bibnamefont {Zhengyou}}, \ and\ \bibinfo {author} {\bibfnamefont {G.}~\bibnamefont {Zhen}},\ }\href {\doibase https://doi.org/10.1016/j.scib.2024.05.003} {\bibfield  {journal} {\bibinfo  {journal} {Sci. Bull.}\ }\textbf {\bibinfo {volume} {69}},\ \bibinfo {pages} {2050} (\bibinfo {year} {2024})}\BibitemShut {NoStop}%
\bibitem [{\citenamefont {Geier}\ \emph {et~al.}(2018)\citenamefont {Geier}, \citenamefont {Trifunovic}, \citenamefont {Hoskam},\ and\ \citenamefont {Brouwer}}]{PhysRevB.97.205135}%
  \BibitemOpen
  \bibfield  {author} {\bibinfo {author} {\bibfnamefont {M.}~\bibnamefont {Geier}}, \bibinfo {author} {\bibfnamefont {L.}~\bibnamefont {Trifunovic}}, \bibinfo {author} {\bibfnamefont {M.}~\bibnamefont {Hoskam}}, \ and\ \bibinfo {author} {\bibfnamefont {P.~W.}\ \bibnamefont {Brouwer}},\ }\href {\doibase 10.1103/PhysRevB.97.205135} {\bibfield  {journal} {\bibinfo  {journal} {Phys. Rev. B}\ }\textbf {\bibinfo {volume} {97}},\ \bibinfo {pages} {205135} (\bibinfo {year} {2018})}\BibitemShut {NoStop}%
\bibitem [{\citenamefont {Khalaf}\ \emph {et~al.}(2021)\citenamefont {Khalaf}, \citenamefont {Benalcazar}, \citenamefont {Hughes},\ and\ \citenamefont {Queiroz}}]{PhysRevResearch.3.013239}%
  \BibitemOpen
  \bibfield  {author} {\bibinfo {author} {\bibfnamefont {E.}~\bibnamefont {Khalaf}}, \bibinfo {author} {\bibfnamefont {W.~A.}\ \bibnamefont {Benalcazar}}, \bibinfo {author} {\bibfnamefont {T.~L.}\ \bibnamefont {Hughes}}, \ and\ \bibinfo {author} {\bibfnamefont {R.}~\bibnamefont {Queiroz}},\ }\href {\doibase 10.1103/PhysRevResearch.3.013239} {\bibfield  {journal} {\bibinfo  {journal} {Phys. Rev. Res.}\ }\textbf {\bibinfo {volume} {3}},\ \bibinfo {pages} {013239} (\bibinfo {year} {2021})}\BibitemShut {NoStop}%
\bibitem [{\citenamefont {Benalcazar}\ \emph {et~al.}(2017)\citenamefont {Benalcazar}, \citenamefont {Bernevig},\ and\ \citenamefont {Hughes}}]{RN12}%
  \BibitemOpen
  \bibfield  {author} {\bibinfo {author} {\bibfnamefont {W.~A.}\ \bibnamefont {Benalcazar}}, \bibinfo {author} {\bibfnamefont {B.~A.}\ \bibnamefont {Bernevig}}, \ and\ \bibinfo {author} {\bibfnamefont {T.~L.}\ \bibnamefont {Hughes}},\ }\href {\doibase doi:10.1126/science.aah6442} {\bibfield  {journal} {\bibinfo  {journal} {Science}\ }\textbf {\bibinfo {volume} {357}},\ \bibinfo {pages} {61} (\bibinfo {year} {2017})}\BibitemShut {NoStop}%
\bibitem [{sup()}]{suppmat}%
  \BibitemOpen
  \href@noop {} {}\bibinfo {note} {Supplementary Material contains information on the derivation of the symmetry operator, the calculation of topological invariant, and the simulation results for $3\times3\times3$ unit cells.}\BibitemShut {Stop}%
\bibitem [{\citenamefont {Zhang}\ \emph {et~al.}(2020{\natexlab{b}})\citenamefont {Zhang}, \citenamefont {Lin}, \citenamefont {Wang}, \citenamefont {Xiong}, \citenamefont {Tian}, \citenamefont {Lu}, \citenamefont {Chen},\ and\ \citenamefont {Jiang}}]{RN13}%
  \BibitemOpen
  \bibfield  {author} {\bibinfo {author} {\bibfnamefont {X.}~\bibnamefont {Zhang}}, \bibinfo {author} {\bibfnamefont {Z.-K.}\ \bibnamefont {Lin}}, \bibinfo {author} {\bibfnamefont {H.-X.}\ \bibnamefont {Wang}}, \bibinfo {author} {\bibfnamefont {Z.}~\bibnamefont {Xiong}}, \bibinfo {author} {\bibfnamefont {Y.}~\bibnamefont {Tian}}, \bibinfo {author} {\bibfnamefont {M.-H.}\ \bibnamefont {Lu}}, \bibinfo {author} {\bibfnamefont {Y.-F.}\ \bibnamefont {Chen}}, \ and\ \bibinfo {author} {\bibfnamefont {J.-H.}\ \bibnamefont {Jiang}},\ }\href {\doibase 10.1038/s41467-019-13861-4} {\bibfield  {journal} {\bibinfo  {journal} {Nat. Commun.}\ }\textbf {\bibinfo {volume} {11}},\ \bibinfo {pages} {65} (\bibinfo {year} {2020}{\natexlab{b}})}\BibitemShut {NoStop}%
\bibitem [{\citenamefont {Wieder}\ \emph {et~al.}(2018)\citenamefont {Wieder}, \citenamefont {Bradlyn}, \citenamefont {Wang}, \citenamefont {Cano}, \citenamefont {Kim}, \citenamefont {Kim}, \citenamefont {Rappe}, \citenamefont {Kane},\ and\ \citenamefont {Bernevig}}]{RN14}%
  \BibitemOpen
  \bibfield  {author} {\bibinfo {author} {\bibfnamefont {B.~J.}\ \bibnamefont {Wieder}}, \bibinfo {author} {\bibfnamefont {B.}~\bibnamefont {Bradlyn}}, \bibinfo {author} {\bibfnamefont {Z.}~\bibnamefont {Wang}}, \bibinfo {author} {\bibfnamefont {J.}~\bibnamefont {Cano}}, \bibinfo {author} {\bibfnamefont {Y.}~\bibnamefont {Kim}}, \bibinfo {author} {\bibfnamefont {H.-S.~D.}\ \bibnamefont {Kim}}, \bibinfo {author} {\bibfnamefont {A.~M.}\ \bibnamefont {Rappe}}, \bibinfo {author} {\bibfnamefont {C.~L.}\ \bibnamefont {Kane}}, \ and\ \bibinfo {author} {\bibfnamefont {B.~A.}\ \bibnamefont {Bernevig}},\ }\href {\doibase doi:10.1126/science.aan2802} {\bibfield  {journal} {\bibinfo  {journal} {Science}\ }\textbf {\bibinfo {volume} {361}},\ \bibinfo {pages} {246} (\bibinfo {year} {2018})}\BibitemShut {NoStop}%
\bibitem [{\citenamefont {Wang}\ \emph {et~al.}(2016)\citenamefont {Wang}, \citenamefont {Alexandradinata}, \citenamefont {Cava},\ and\ \citenamefont {Bernevig}}]{RN15}%
  \BibitemOpen
  \bibfield  {author} {\bibinfo {author} {\bibfnamefont {Z.}~\bibnamefont {Wang}}, \bibinfo {author} {\bibfnamefont {A.}~\bibnamefont {Alexandradinata}}, \bibinfo {author} {\bibfnamefont {R.~J.}\ \bibnamefont {Cava}}, \ and\ \bibinfo {author} {\bibfnamefont {B.~A.}\ \bibnamefont {Bernevig}},\ }\href {\doibase 10.1038/nature17410} {\bibfield  {journal} {\bibinfo  {journal} {Nature}\ }\textbf {\bibinfo {volume} {532}},\ \bibinfo {pages} {189} (\bibinfo {year} {2016})}\BibitemShut {NoStop}%
\bibitem [{\citenamefont {Yang}\ \emph {et~al.}(2022)\citenamefont {Yang}, \citenamefont {Po}, \citenamefont {Liu}, \citenamefont {Joannopoulos}, \citenamefont {Fu},\ and\ \citenamefont {Solja\ifmmode \check{c}\else \v{c}\fi{}i\ifmmode~\acute{c}\else \'{c}\fi{}}}]{PhysRevB.106.L161108}%
  \BibitemOpen
  \bibfield  {author} {\bibinfo {author} {\bibfnamefont {Y.}~\bibnamefont {Yang}}, \bibinfo {author} {\bibfnamefont {H.~C.}\ \bibnamefont {Po}}, \bibinfo {author} {\bibfnamefont {V.}~\bibnamefont {Liu}}, \bibinfo {author} {\bibfnamefont {J.~D.}\ \bibnamefont {Joannopoulos}}, \bibinfo {author} {\bibfnamefont {L.}~\bibnamefont {Fu}}, \ and\ \bibinfo {author} {\bibfnamefont {M.}~\bibnamefont {Solja\ifmmode \check{c}\else \v{c}\fi{}i\ifmmode~\acute{c}\else \'{c}\fi{}}},\ }\href {\doibase 10.1103/PhysRevB.106.L161108} {\bibfield  {journal} {\bibinfo  {journal} {Phys. Rev. B}\ }\textbf {\bibinfo {volume} {106}},\ \bibinfo {pages} {L161108} (\bibinfo {year} {2022})}\BibitemShut {NoStop}%
\bibitem [{\citenamefont {Shuo}\ \emph {et~al.}(2019)\citenamefont {Shuo}, \citenamefont {Wenlong}, \citenamefont {Qian}, \citenamefont {Shaojie}, \citenamefont {Lei}, \citenamefont {Changxu}, \citenamefont {Yuan~Jiang}, \citenamefont {Tie~Jun},\ and\ \citenamefont {Shuang}}]{RN9}%
  \BibitemOpen
  \bibfield  {author} {\bibinfo {author} {\bibfnamefont {L.}~\bibnamefont {Shuo}}, \bibinfo {author} {\bibfnamefont {G.}~\bibnamefont {Wenlong}}, \bibinfo {author} {\bibfnamefont {Z.}~\bibnamefont {Qian}}, \bibinfo {author} {\bibfnamefont {M.}~\bibnamefont {Shaojie}}, \bibinfo {author} {\bibfnamefont {Z.}~\bibnamefont {Lei}}, \bibinfo {author} {\bibfnamefont {L.}~\bibnamefont {Changxu}}, \bibinfo {author} {\bibfnamefont {X.}~\bibnamefont {Yuan~Jiang}}, \bibinfo {author} {\bibfnamefont {C.}~\bibnamefont {Tie~Jun}}, \ and\ \bibinfo {author} {\bibfnamefont {Z.}~\bibnamefont {Shuang}},\ }\href {\doibase 10.34133/2019/8609875} {\bibfield  {journal} {\bibinfo  {journal} {Research}\ }\textbf {\bibinfo {volume} {2019}} (\bibinfo {year} {2019}),\ 10.34133/2019/8609875}\BibitemShut {NoStop}%
\bibitem [{\citenamefont {R\"ontgen}\ \emph {et~al.}(2024)\citenamefont {R\"ontgen}, \citenamefont {Chen}, \citenamefont {Gao}, \citenamefont {Pyzh}, \citenamefont {Schmelcher}, \citenamefont {Pagneux}, \citenamefont {Achilleos},\ and\ \citenamefont {Coutant}}]{PhysRevB.110.035106}%
  \BibitemOpen
  \bibfield  {author} {\bibinfo {author} {\bibfnamefont {M.}~\bibnamefont {R\"ontgen}}, \bibinfo {author} {\bibfnamefont {X.}~\bibnamefont {Chen}}, \bibinfo {author} {\bibfnamefont {W.}~\bibnamefont {Gao}}, \bibinfo {author} {\bibfnamefont {M.}~\bibnamefont {Pyzh}}, \bibinfo {author} {\bibfnamefont {P.}~\bibnamefont {Schmelcher}}, \bibinfo {author} {\bibfnamefont {V.}~\bibnamefont {Pagneux}}, \bibinfo {author} {\bibfnamefont {V.}~\bibnamefont {Achilleos}}, \ and\ \bibinfo {author} {\bibfnamefont {A.}~\bibnamefont {Coutant}},\ }\href {\doibase 10.1103/PhysRevB.110.035106} {\bibfield  {journal} {\bibinfo  {journal} {Phys. Rev. B}\ }\textbf {\bibinfo {volume} {110}},\ \bibinfo {pages} {035106} (\bibinfo {year} {2024})}\BibitemShut {NoStop}%
\bibitem [{\citenamefont {Chen}(2003)}]{chen2003circuits}%
  \BibitemOpen
  \bibfield  {author} {\bibinfo {author} {\bibfnamefont {W.-K.}\ \bibnamefont {Chen}},\ }\href@noop {} {\emph {\bibinfo {title} {The Circuits and Filters Handbook}}}\ (\bibinfo  {publisher} {CRC Press},\ \bibinfo {year} {2003})\ pp.\ \bibinfo {pages} {396--397}\BibitemShut {NoStop}%
\bibitem [{\citenamefont {Shao}\ \emph {et~al.}(2023)\citenamefont {Shao}, \citenamefont {Chen}, \citenamefont {Wang}, \citenamefont {Yang},\ and\ \citenamefont {Zhao}}]{PhysRevB.108.205126}%
  \BibitemOpen
  \bibfield  {author} {\bibinfo {author} {\bibfnamefont {L.}~\bibnamefont {Shao}}, \bibinfo {author} {\bibfnamefont {Z.}~\bibnamefont {Chen}}, \bibinfo {author} {\bibfnamefont {K.}~\bibnamefont {Wang}}, \bibinfo {author} {\bibfnamefont {S.~A.}\ \bibnamefont {Yang}}, \ and\ \bibinfo {author} {\bibfnamefont {Y.}~\bibnamefont {Zhao}},\ }\href {\doibase 10.1103/PhysRevB.108.205126} {\bibfield  {journal} {\bibinfo  {journal} {Phys. Rev. B}\ }\textbf {\bibinfo {volume} {108}},\ \bibinfo {pages} {205126} (\bibinfo {year} {2023})}\BibitemShut {NoStop}%
\end{thebibliography}
\end{document}